       \documentclass[final,1p,times]{elsarticle}
\usepackage[utf8]{inputenc}  
\usepackage{graphicx,color} 
\usepackage{mathrsfs}
\usepackage{enumitem}
\usepackage{bm}
\usepackage{array}
\usepackage{amsmath}
\usepackage{amssymb}
\usepackage{amsfonts}
 \usepackage{amsbsy}
 \usepackage{bbold}
\usepackage{bbm}
 \usepackage{ulem}
 \usepackage{siunitx}
\usepackage{booktabs,cancel,caption,cleveref,colortbl,csquotes}
\usepackage{helvet,mathpazo,multirow,listings,pgfplots,xcolor}
\usepackage{textcomp}
\usepackage{lineno}
\usepackage{listings}
\usepackage{xfrac}
\usepackage{fleqn}
\usepackage{verbatimbox}
\usepackage{textcomp}
\usepackage{float}
\usepackage{caption}
\usepackage{url}
\usepackage{mathtools}
\usepackage{lipsum}

\usepackage{lineno}

\newcounter{bla}

\journal{Computer Physics Communications}

\begin{document}


\begin{frontmatter}



\title{SWANLOP : Scattering waves off nonlocal optical potentials in
       the presence of Coulomb interactions}


  \author[a,b]{H.~F.~Arellano \corref{author}}
  \author[b]{G.~Blanchon}

\cortext[author] 
{Corresponding author.\\\textit{E-mail address:} arellano@dfi.uchile.cl}
\address[a]{Department of Physics - FCFM, 
  University of Chile, Av. Blanco Encalada 2008, Santiago, Chile}
\address[b]{CEA,DAM,DIF F-91297 Arpajon, France}

\begin{abstract}
  We introduce the package {\small SWANLOP} 
  to calculate scattering waves and corresponding observables 
for nucleon elastic collisions off spin-zero nuclei.
The code is capable of handling local and nonlocal optical 
potentials superposed to long-range Coulomb interaction. 
Solutions to the implied Schr\"odinger integro-differential
equation are obtained by solving 
an integral equation of Lippmann-Schwinger type for the scattering 
wavefunctions, $\psi=\phi_{_C} + {G}_{_C} {U}_{_S}\psi$,
providing and exact treatment
to the Coulomb force [Phys. Lett. B 789, 256 (2019)]. 
The package has been developed to handle potentials 
either in momentum or coordinate representations,
providing flexible options under each of them.
The code is fully self-contained,
being dimensioned to handle any $A\!\geq\!4$ target
for nucleon beam energies of up to \num{1.1}~GeV.
Accuracy and benchmark applications are presented and discussed.
\end{abstract}


\end{frontmatter}



{\bf Program summary}

\begin{small}
\noindent
{\it Program title:}            
  SWANLOP
\\
{\it Catalog identifier:}                                             
\\
{\it Program summary URL:}                                              
\\
{\it Program obtainable from:}  
  CPC Program Library, Queen's University, Belfast, N. Ireland    
\\
{\it Licensing provisions:}     
  GNU General Public License, Version 2   
\\
{\it No. of lines in distributed program, including test data, etc:}    
\\
{\it No. of bytes in distributed program, including test data, etc:}    
\\
{\it Distribution format:}      tar.gz
\\ 
{\it Programming language:}     FORTRAN-90
\\
{\it Computer:}
\\
{\it Operating system:}         LINUX, Mac OS
\\
{\it RAM:}                      Memory usage depends on ...             
\\
{\it Classification:}
\\
{\it Nature of problem:}        
Optical model potentials constitute a valuable tool to investigate 
the physics involved in nuclear collisions and reactions 
with nucleonic probes.
As such, it becomes essential to obtain accurate results for its 
associated scattering observables and corresponding scattering waves.
An important feature of optical potentials is their nonlocal nature,
arising from the fermionic nature of the $(\!A\!+\!1)$--nucleon
problem together with the fact that effective nucleon-nucleon 
(\textit{NN}) interactions are nonlocal as well.
The superposition of Coulomb interaction to these nonlocal potentials
poses non-trivial difficulties to obtain scattering waves and
observables in collision processes.
\\
{\it Solution method:}
The code performs the calculation of scattering waves associated 
to nonlocal potentials in the presence of the long-range Coulomb 
interactions, solving 
a Lippmann-Schwinger type
integral equation for the scattering wavefunction.
The potential can be given either in coordinate or momentum space.
Phase-shifts and associated elastic scattering observables
are extracted from the asymptotic behavior of the solution.
\\
{\it Running time: 
    The code takes from 1~s, 
    in the case of low-energy nucleon scattering off light targets,
    up to \num{100}~s for 1-GeV nucleons off heavy targets,
    using conventional 2.6~GHz laptop computer.
}
\\
\end{small}

\section{Introduction}
\label{sec:introduction}
Current developments in theoretical nuclear research 
have set their focus on the development
and calculation of non-Hermitian, nonlocal and energy-dependent 
optical potentials to describe the interaction of nucleonic 
probes with nuclei.
Important achievements in these efforts have been 
\textit{ab-initio} approaches reported in Refs.
\cite{Rotureau2017,Idini2019}, 
the construction of potentials based on energy density 
functionals \cite{Mizuyama2012c,Hao2015,Blanchon2015}, 
the calculation of $g$-matrix based optical potentials
\cite{Arellano1995,Dupuis2006},
in addition to $t$-matrix based optical models
\cite{Arellano1990,Elster1990,Crespo1990,Weppner1998,Vorabbi2016,
Burrows2019}.
With these advances in mind, the accurate treatment of 
intrinsic nonlocalities of these potentials in collision
processes becomes crucial in order to investigate objectively
their physical implications.

In the presence of nonlocal couplings between the projectile and target,
Schr\"odinger equation for scattering waves 
becomes an integro-differential 
equation in coordinate space. 
Furthermore, the superposition of Coulomb interaction 
to these nonlocal potentials poses non-trivial difficulties 
to obtain scattering waves and observables in collision processes.
In this work we introduce the package 
{\small SWANLOP} 
aimed to perform such calculations by solving 
an integral equation for the scattering wavefunction
of Lippmann-Schwinger type.
The solution to the problem is formally exact as reported 
in Ref.~\cite{Arellano2019},
where the scattering wave gets expressed in terms of known quantities.
Optical potentials in momentum representation are treated as well.
The resulting scattering waves can further be used
in distorted wave Born approximations. 
The acronym {\small SWANLOP} 
stands for Scattering WAves off NonLocal Optical Potentials.

Several methods have been reported to solve the scattering
problem under nonlocal potentials.
Early solutions to this problem were proposed 
by Perey and Buck (PB)\cite{Perey1962}, 
where the separable structure of the potential is used to isolate 
the role of the nonlocal factor,
reducing the integro-differential Schr\"odinger equation into a
second-order differential equation with a local coupling.
A known disadvantage of this approach is that the resulting 
scattering waves differ from the exact ones, 
distortion coined as Perey effect being characterized 
by a Perey correction factor \cite{Titus2014}.

Other solutions to Schr\"odinger's integro-differential equation 
follow iterative procedures \cite{Perey1962,Titus2014,Titus2016}. 
In these approaches Schr\"odinger's differential equation 
is integrated with a non-homogeneous term consisting of 
the projection of the nonlocal coupling  
onto an intermediate solution. 
Iterations start with a given seed for the scattering wave,
solving Schr\"odinger equation in the presence of a 
non-homogeneous term.
A drawback of this method is that prior knowledge
of the solution is needed for efficient convergence,
though there is no theoretical assurance
to converge to the actual solution. 

In the case of Ref. \cite{Upadhyay2018}, a mean-value 
approximation is applied for the coupling of the nonlocal
term with the scattering wave, reducing the problem to a second-order 
homogeneous differential equation. 
This method is restricted to neutron collisions.
Quite recently another approach has been proposed to deal with 
nonlocal potentials \cite{Upadhyay2018b}, 
resorting to a Taylor approximation for the radial wave function.
The method assumes that nonlocality is dominant around 
the diagonal in coordinate space, feature which is non universal
as observed in coordinate-space representations of
potentials originally calculated in momentum space~\cite{Arellano2018}.

Solutions to the scattering problem in momentum space
have also been investigated 
\cite{Picklesimer1984,Arellano1990,Elster1990,Crespo1990,
Chinn1991,Lu1994,Upadhyay2014}. 
While an appealing advantage of momentum-space approaches is that 
nonlocalities are naturally accounted for, 
one of its limitations when long-range Coulomb interactions are 
included is that the associated scattering waves are not readily 
available. Not only that, but the long range of the Coulomb interaction 
results in a $\sim$$1/q^{2}$ singularity, feature 
that has led to the use of specific procedures at the moment of
calculating scattering amplitudes.
An exact solution addressing this singularity has been proposed 
by Vincent and Phatak by means of a 
cut-off technique to the Coulomb long-range 
tail \cite{Vincent1974}. 
In this way it is possible to obtain the exact 
(on-shell) scattering amplitude from the solution for the
screened potential.
This approach has been applied to proton-nucleus 
(\textit{pA}) scattering at 
intermediate energies \cite{Arellano1990}, where its accuracy is 
significantly improved after a detailed multipole treatment 
of the charge form factor convoluted with a sharp cut-off 
point Coulomb potential,
as discussed by Einsenstein and Tabakin~\cite{Eisenstein1982}.

In works by Alt \textsl{et al.}~\cite{Alt1978,Alt1980}
the Coulomb long-range potential is screened with the use of smooth
radial form factors, resulting in finite-range interactions.
The associated scattering matrix can then be calculated using standard 
techniques.
The zero-screening limit is obtained by increasing the range
$R$ of the form factor in conjunction with the use of 
renormalization factors. 
This method has been refined by Deltuva and 
collaborators~\cite{Deltuva2005a,Deltuva2005b} 
in studies of three-nucleon breakup reactions in momentum space.
In their work exponential screening form factors 
of type $\sim\!\exp[-(r/R)^4]$ are used.

  Studies pursued by 
  Elster and collaborators~\cite{Elster1990r,Chinn1991,Elster1993}
  have addressed the \textit{pA} scattering problem without 
  resorting to screening techniques.
  Here the full \textit{pA} interaction is re-expressed as the sum
  of a point Coulomb term and short-range residuum.
  The use of two-potential formalism enables to express the
  scattering amplitude as the sum of two terms.
  A residual Coulomb-modified transition matrix is
  obtained solving a Lippmann-Schwinger equation for
  a modified potential which includes Coulomb distortions.
  Calculated scattering observables for \textit{pA} 
  scattering are accurate even for 500-MeV protons off heavy targets.

Another method to calculate waves off nonlocal potentials 
in the presence of long-range Coulomb 
interaction is that of Refs.~\cite{Kim1990,Kim1992}, 
where Lanczos technique is used to solve integral 
equations derived from the nonlocal Schr\"odinger equation. 
Later on, in Refs.~\cite{Hagen2012,Michel2011} 
a numerical treatment to this problem is presented with
the use of Berggren basis, where an 
off-diagonal approximation is used to control the Coulomb 
singularity along the diagonal in momentum space. 
Applications of this approach have been reported for low 
energies and intermediate-mass targets.

Quite recently the package {\small SIDES} 
(Schr\"odinger Integro-Differential Equation Solver)
has been introduced~\cite{Blanchon2020}, featuring an
exact treatment of the long-range Coulomb interaction. 
The approach is based on finite difference 
techniques~\cite{Raynal1998,Gibbs2006}, 
where the integro-differential equation
in coordinate space is reduced to
a matrix equation for the wavefunction. 
This approach contrasts with the method we use in 
{\small SWANLOP},
where wavefunctions are obtained from 
an
integral equation for the wavefunction, including Coulomb interactions.
Additionally, {\small SWANLOP} features the possibility
of working with potentials given in momentum space.

This paper is organized as follows. 
In Sec.~\ref{sec:framework} we lay out the framework and present 
a formal solution to the scattering problem with nonlocal 
potentials in the presence of Coulomb interactions. 
We also establish contact with potentials represented in momentum
space, providing transformation into coordinate representation, 
to obtain exact scattering observables in the
presence of Coulomb interaction.
In Sec.~\ref{sec:package} we describe the 
{\small SWANLOP} package,
its I/O structure, main options and execution of the code.
In Sec.~\ref{sec:benchmark} we study the accuracy of 
{\small SWANLOP}
by comparing with analytic solutions, exploring convergence 
on integration step length and comparing results with
the recently released package {\small SIDES}~\cite{Blanchon2020}.
Additionally, we discuss CPU run-time performance of the code.
In Sec.~\ref{sec:summary} we present a summary and conclusions of
this work.

\section{Framework}
\label{sec:framework}
In this section we layout key equations needed to describe
\textit{NA} collisions under nonlocal potentials 
(in coordinate space) superposed to Coulomb forces. 
We present the solution to the scattering problem and 
make contact with potentials expressed in momentum representation.
For details on the derivation of the solution we refer 
the reader to Ref.~\cite{Arellano2019}.

Consider a proton of mass $m$ 
with kinetic energy $E_{lab}$ in the laboratory reference frame,
colliding a 
 spin-zero
nucleus of mass $M$ and charge $Ze$ at rest. 
Let $U$ the full interaction between them,
being comprised of a pure hadronic contribution $U_H$ and Coulomb 
interaction $V_{_C}$ due to the distributed charge in the nucleus.
The hadronic part is short-range so that the total interaction 
can be cast as the sum of point-Coulomb and short-range terms,
\begin{equation}
  \label{us}
U({\bm r}',{\bm r}) = 
U^{(s)}({\bm r}',{\bm r}) + \frac{\beta}{r}\,\delta({\bm r}'-{\bm r})\;,
\end{equation}
with $\beta=Ze^2$.
Here $U^{(s)}\!=\!U_H\!+\!V_{_C}\!-\!\beta\,\delta({\bm r}'-{\bm r})/r$,
which vanishes rapidly away from the nucleus.
  In the case of neutron scattering both $\beta$ and $V_{_C}$ vanish,
so that $U^{(s)}\!=\!U_H$, being this a particular case in the
discussion that follows.

With the above Schr\"odinger's equation for scattering waves
in the center-of-momentum reference frame reads
\begin{equation}
\label{schr}
-\nabla^2 \psi_{\bm k} ({\bm r}) + 
\frac{2\mu}{\hbar^2} 
\int d{\bm r'} U({\bm r},{\bm r'})\psi_{\bm k}({\bm r'})=
k^2 \psi_{\bm k}({\bm r}) \; ,
\end{equation}
where $\mu$ denotes the \textit{NA} reduced mass and
$k$ the asymptotic relative momentum in the \textit{NA}
center-of-momentum reference frame. 
We omit spin and isospin variables for simplicity in the notation.
Consistent with spin$-\sfrac{1}{2}$ nucleons colliding
a spherical target we expand
\begin{equation}
\label{expansion}
\psi_{\bm k} ({\bm r}) =
  \sqrt{\frac{2}{\pi}}\;
  \sum_{jlm_j} i^l 
  {\cal Y}_{jl1/2}^{m_j}({\hat{\bm r}})
e^{i\sigma_l}
\frac{u_{jl}(r)}{r} 
  {\cal Y}_{jl1/2}^{m_j\dagger}({\hat{\bm k}}) \; .
\end{equation}
Here $u_{jl}(r)$ denotes the radial wavefunction
and $\sigma_l$ the Coulomb phase-shift for partial wave $l$. 
Furthermore,
$\mathcal{Y}_{jls}^{m_j}$ stand for spin 
$s\!=\!\sfrac12$ spherical vectors
\begin{equation}
  \mathcal{Y}_{jls}^{m_j}(\hat{\bm k}) = \sum_{m m_s}
  Y_{l}^{m}(\hat{\bm k})\,|s m_s\rangle\langle ls\,m m_s|jm_j\rangle\,.
\end{equation}
The normalization adopted in Eq.~\eqref{expansion} for 
$\psi_{\bm k} ({\bm r})$ is such that it reduces to normalized 
plane waves 
$\sim\!e^{i{\bm k}\cdot{\bm r}}/(2\pi)^{3/2}$,
when interactions are fully suppressed.

Replacing $\psi_{\bm k} ({\bm r})$ from Eq.~\eqref{expansion}
into Eq.~\eqref{schr}, 
following standard procedures we get
\begin{equation}
  \label{sch}
\left [
\frac{1}{r}
\left ( 
\frac{d^2}{dr^2}
\right ) r - \frac{l(l+1)}{r^2} + k^2
\right ] \frac{u_{jl}(r)}{r} = 
\frac{2\mu}{\hbar^2}
  \int_{0}^{\infty} r'\,dr'
U_{jl}(r,r')
u_{jl}(r') \; ,
\end{equation}
where the multipoles $U_{jl}(r',r)$ of the interaction are obtained from
\begin{equation}
  \label{Rmultip}
U_{jl}(r',r) = \iint d\hat{\bm r}\,d\hat{\bm r}' 
  {\cal Y}_{jl1/2}^{m_j \dagger}({\hat{\bm r}}') 
U({\bm r'},{\bm r})
  {\cal Y}_{jl1/2}^{m_j}(\hat{\bm r}) \;.
\end{equation}
Making explicit the separation of the interaction
into a pointlike source and finite-range remaining
\begin{equation}
U_{jl}(r',r) \equiv U_{jl}^{(s)}(r',r) + \frac{\beta}{r^3}\delta(r'-r) \; ,
\end{equation}
we obtain
\begin{equation}
\label{schr2}
\left [
\frac{d^2}{dr^2} - \frac{l(l+1)}{r^2} -\frac{2k\eta}{r}+ k^2
\right ] u_{jl}(r) 
 =
\frac{2\mu}{\hbar^2} 
  \int_{0}^{\infty} dr' rU_{jl}^{(s)}(r,r')r' u_{jl}(r') \; ,
\end{equation}
with $\eta$ the Sommerfeld parameter given by 
$\eta\!=\!\mu\beta/\hbar^2 k$.
Following Ref. \cite{Arellano2019},
a formal solution to this equation is expressed as the 
superposition of homogeneous and particular solutions in the form
\begin{equation}
\label{solution}
u_{jl}(r) = 
\frac{1}{k} F_l(\eta,kr) + 
\frac{2\mu}{\hbar^2}
\iint dr'dr'' 
 G_l^{c(+)}(r,r';k) \left [ 
r'U^{(s)}_{jl}(r',r'') r'' \right ] u_{jl}(r'') \; ,
\end{equation}
with the Coulomb propagator
\begin{equation}
\label{propagator}
G_l^{c(+)}(r,r';k) 
 = 
-\frac{i}{k}
F_l(\eta,kr_{<})
  \left [
    F_{l}(\eta,kr_{>})
 -i G_{l}(\eta,kr_{>})
  \right ]
    \;,
\end{equation}
where 
$r_{<}\!=\!\textrm{min}(r,r')$, and
$r_{>}\!=\!\textrm{max}(r,r')$.
In the above $F_{l}$ and $G_{l}$ denote regular and irregular 
Coulomb functions~\cite{Abramowitz1965} under the phase convention
\begin{align}
\label{ric2}
  F_l(\eta,z) &\xrightarrow[\;z\to\infty]{}
      \sin(z - \eta\ln 2z -l\pi/2 + \sigma_l)\;,
\nonumber \\
  G_l(\eta,z) &\xrightarrow[\;z\to\infty]{}
      \cos(z - \eta\ln 2z -l\pi/2 + \sigma_l)\;.
\end{align}

Note that the Coulomb propagator expressed by Eq.~\eqref{propagator} 
is non-singular, being a continuous function of $r$ and $r'$. 
The spatial gradient of $G_l^{c(+)}(r',r;k)$ is discontinuous 
at the diagonal $r\!=\!r'$, feature that poses no particular drawback. 
Furthermore, Eq.~\eqref{solution} takes the form of 
an
integral equation for scattering waves in the presence 
of Coulomb interaction, 
which we recast as
\begin{equation}
\label{ekernel}
  \int_{0}^{\infty} dr''
  \left [
  \delta(r-r'') -
  K_{jl}(r,r'')
  \right ] 
  u_{jl}(r'') = 
  \textstyle{\frac{1}{k}} F_l(\eta,kr)\;,
\end{equation}
where the kernel $K_{jl}$ is given by
\begin{equation}
\label{kernel}
K_{jl}(r,r'') =
\frac{2\mu}{\hbar^2}
  \int_{0}^{\infty} dr' 
 G_l^{c(+)}(r,r';k) 
  \left [ 
r'U^{(s)}_{jl}(r',r'') r'' 
  \right ] \; .
\end{equation}
Note that Eq.~\eqref{ekernel} enables to obtain the actual 
scattering wavefunction by means of direct matrix inversion. 

The solution for $u_{jl}$ from Eq. \eqref{ekernel} enables the 
calculation of the scattering amplitude, which follows from
the asymptotic form of Eq.~\eqref{solution}, where $r$ is taken 
far away from the scattering center. 
In this limit we have
\begin{equation}
\label{faraway2}
G_l^{c(+)}(r,r';k) 
  \xrightarrow[r\gg r']{}
 -
\frac{i}{k} 
F_l(\eta,kr')
  \left [
    F_{l}(\eta,kr)
 -i G_{l}(\eta,kr)
  \right ] \;,
\end{equation}
which once replaced in Eq.~\eqref{solution} for $u_{jl}$ yields
\begin{equation}
\label{match}
k\,u_{jl}(r)
  \xrightarrow[r\to\infty]{}
F_l(\eta,kr) + \Delta_{jl}  
\left [ F_l(\eta,kr) - iG_l(\eta,kr) \right ] ,
\end{equation}
with  
\begin{equation}
\Delta_{jl} = - 
\frac{2\mu i}{\hbar^2} 
\iint r'dr'\,r'' dr''
 F_l(\eta,kr') U^{(s)}_{jl}(r',r'')  u_{jl}(r'') \; .
\end{equation}
These last two relations allow independent ways to obtain $\Delta_{jl}$. 
The latter involves direct integration of the wavefunction whereas
the former evaluates asymptotically the ratio
\begin{equation}
\label{asymp}
\Delta_{jl} = \frac{ku_{jl}(r) - F_l(\eta,kr)}
                   {F_l(\eta,kr) - iG_l(\eta,kr) }\;,
\end{equation}
for sufficiently large $r$.
These last two equivalent forms for $\Delta_{jl}$ are
useful for consistency checks.
Once $\Delta_{jl}$ is obtained, the scattering amplitude $f_{jl}$ 
and short-range phase shift $\bar\delta_{jl}$ follow from
\begin{equation}
\Delta_{jl} =
 ikf_{jl} =
\textstyle{\frac12}\left (e^{ 2i\bar\delta_{jl}} - 1 \right ).
\end{equation}
Later on it will be useful to refer to the $S$ matrix associated
to $\bar\delta_{jl}$, defined by
\begin{equation}
  \label{sbar}
  \bar S_{jl}=e^{2i\bar\delta_{jl}}\;.
\end{equation}

The numerical implementation of Eq. \eqref{ekernel} follows from
the discretization of $r$ (and $r''$) over an $N$-point
uniform mesh up to $r\!=\!R_{max}$.
The $n$-th element of this array is given by $r_n\!=\!n\,h$, 
with $h\!=\!R_{max}/N$.
We find trapezoidal rule adequate to evaluate the integrals.
The kernel in Eq.~\eqref{kernel}, function of $r$ and $r'$, 
becomes a finite $N\!\times\! N$ 
matrix which we denote by $\mathbb{K}$. 
This kernel is fully determined by the matrix elements of the
potential and free Coulomb functions, all of them known quantities.
The solution to Eq.~\eqref{ekernel} takes the form
\begin{equation}
\label{matrixeq}
\textrm{\bf u} =  (1 - \mathbb{K})^{-1}\textrm{\bf u}_0\;,
\end{equation}
where $\textrm{\bf u}_0$ represents the unperturbed wave $F_l(\eta,kr)/k$, 
and $\textrm{\bf u}$ denotes the scattering wave over the discrete mesh. 
In this way the scattering wavefunction is directly determined
by inverting a known matrix, which is then multiplied to a known vector.
There is no need to introduce normalization constants
nor the calculation of derivatives to match 
asymptotic behaviors~\cite{Arellano2019}.

\subsection{Potential in momentum space}
This section is aimed to provide explicit relationships between
potentials represented in momentum space, with their coordinate
space counterparts $U_{jl}(r',r)$ in Eq.~\eqref{kernel} for
the kernel.
As already mentioned,
microscopic optical model potentials in momentum space
have the appealing feature of incorporating in a natural
way intrinsic nonlocalities in $(\!A\!+\!1)$--nucleon systems.  
Calculations of these potentials are performed in momentum space
by folding the ground-state mixed density with an effective
interaction. At intermediate nucleon energies, the \textit{NN}
effective interaction can be taken as the free $t$ matrix
\cite{Arellano1990,Elster1990,Crespo1990,Weppner1998,Vorabbi2016}.
At lower energies the use of the density-dependent 
Brueckner-Bethe-Goldstone $g$ matrix becomes suitable
\cite{Arellano1995,Aguayo2008}.
In all these approaches the optical potential for 
\textit{NA} elastic scattering, $\tilde U({\bm k}',{\bm k};E)$,
can be cast in the form
\begin{equation}
  \label{ukk}
  \tilde U({\bm k'},{\bm k}) =
  \tilde U_{0}({\bm k'},{\bm k}) + 
  i{\bm\sigma}\cdot\hat{\bm n}\,
  \tilde U_{1}({\bm k'},{\bm k})\;,
\end{equation}
with $\hat{\bm n}$ the unit vector 
perpendicular to the scattering plane given by
\begin{equation}
  \hat{\bm n}=\frac{\bm k'\times\bm k}{|\bm k'\times\bm k|}\,,
\end{equation}
and ${\bm\sigma}$ the spin of the projectile.
Here $\tilde U_{0}$ and $\tilde U_{1}$ represent central 
and spin-orbit components of the potential, 
which we assume calculated over a grid of relative
momenta, $k$ and $k'$, and angles between ${\bm k}$ and ${\bm k}'$
expressed by $u\!=\!\hat{\bm k}\cdot\hat{\bm k}'$.
With these considerations in mind, we express
$\tilde U_{0}\!=\!\tilde U_{0}(k',k;u)$, and
$\tilde U_{1}\!=\!\tilde U_{1}(k',k;u)$.
In what follows we seek the relationship between these two
terms and $U_{jl}(r',r)$ needed in Eq.~\eqref{sch}
to obtain its associated scattering waves.

Consistent with Eq. \eqref{Rmultip}, let us expand
\begin{equation}
  \label{Uyy}
  \tilde U( \bm k',\bm k)=
  \sum_{jm_jl}
  \mathcal{Y}_{jls}^{m_j}(\hat{\bm k}')
  \tilde U_{jl}(k',k)
  \mathcal{Y}_{jls}^{m_j\dagger}(\hat{\bm k}),
\end{equation}
Let us also consider the identity
\begin{equation}
  \sum_{m_j=-j}^{j}
  \mathcal{Y}_{jls}^{m_j}(\hat{\bm k}')
  \mathcal{Y}_{jls}^{m_j\dagger}(\hat{\bm k}) =
  \frac{(2j+1)}{8\pi}
  \left [ 
  \rule[18pt]{0pt}{0pt}
  P_l(u)\mathbb{1}_\sigma + \right .
  \left . i{\bm\sigma}\cdot\hat{\bm n}\,
    \frac{\langle {\bm \ell}\cdot{\bm \sigma}\rangle_{jl}}{l(l+1)}
    \,P_{l}^{1}(u)
  \right ],
  \label{yy}
\end{equation}
with $P_l^1(u)=\sqrt{1-u^2}\,dP_l(u)/du$, 
the associated Legendre polynomia.
Additionally,
$  \langle{\bm\ell}\cdot{\bm\sigma}\rangle_{jl} = 
  j(j+1)-l(l+1)-\sfrac{3}{4}$.
Combining Eqs. \eqref{ukk}, \eqref{Uyy} and \eqref{yy} we identify
\begin{subequations}
\begin{align}
  \label{u0}
  \tilde U_{0}(k',k;u) &= 
  \sum_{jl}
  \frac{(2j+1)}{8\pi}
  \tilde U_{jl}(k',k)\,
    P_l(u)
  \\
  \label{u1}
  \tilde U_{1}(k',k;u) &=
  \sum_{jl}
    \left [
  \frac{(2j+1)\langle {\bm \ell}\cdot{\bm \sigma}\rangle_{jl}}
       {8\pi\; l(l+1)}
    \right ]
   \tilde U_{jl}(k',k)\,
    \,P_{l}^{1}(u) \;.
\end{align}
\end{subequations}
Using orthogonality of Legendre polynomia we get
\begin{subequations}
\begin{align}
  \sum_{j=l-1/2}^{l+1/2} 
  \frac{(2j+1)}{(2l+1)}
  \tilde U_{jl}(k',k) &= 4\pi
\int_{-1}^{1} \tilde U_{0}(k',k;u)\,P_l(u)\,du \\
  \sum_{j=l-1/2}^{l+1/2} 
  \frac{(2j+1)\langle {\bm \ell}\cdot{\bm \sigma}\rangle_{jl}}
  {(2l+1)}
  \tilde U_{jl}(k',k) &= 4\pi
  \int_{-1}^{1} \tilde U_{1}(k',k;u)\,P_{l}^{1}(u)\,du \;.
\end{align}
\end{subequations}
From these two equations we obtain
\begin{equation}
  \label{Ujl}
  U_{jl}(k',k)=
  M_{l}^{(0)}(k',k) + 
  \frac{\langle {\bm \ell}\cdot{\bm \sigma}\rangle_{jl}}{l(l+1)}\,
  M_{l}^{(1)}(k',k)\;,
\end{equation}
where 
\begin{subequations}
\begin{align}
  \label{m0}
  M_{l}^{(0)}(k',k) &= 
  2\pi \int_{-1}^{1} \tilde U_{0}(k',k;u)\,P_l(u)\,du \\
  \label{m1}
  M_{l}^{(1)}(k',k) &= 
  2\pi \int_{-1}^{1} \tilde U_{1}(k',k;u)\,P_{l}^{1}(u)\,du\;.
\end{align}
\end{subequations}

With $\tilde U_{jl}(k',k)$ given by Eq. \eqref{Ujl}
we proceed to obtain its coordinate-space counterpart,
which we expand as
\begin{equation}
  \label{Urr}
  U( \bm r',\bm r)=
  \sum_{jm_jl}
  \mathcal{Y}_{jls}^{m_j}(\hat{\bm r}')
  U_{jl}(r',r)
  \mathcal{Y}_{jls}^{m_j\dagger}(\hat{\bm r}).
\end{equation}
Using normalized plane waves
\begin{equation}
  \label{pw}
  \langle {\bm r}|{\bm k}\rangle =
  \frac{e^{i{\bm k}\cdot{\bm r}}} {(2\pi)^{3/2}}\,
  =
  \sqrt{\frac{2}{\pi}}
  \sum_{lm} 
  Y_{l}^{m}(\hat{\bm r}) 
  i^{l} j_{l}(kr)
  Y_{l}^{m*}(\hat{\bm k}) ,
\end{equation}
we evaluate
\begin{equation}
  \label{qtor}
U({\bm r}',{\bm r}) = \iint d{\bm k}'d{\bm k}
\langle {\bm r}'|{\bm k}'\rangle\,
\tilde U({\bm k'},{\bm k}) \,
\langle {\bm k}|{\bm r}\rangle,
\end{equation}
to obtain
\begin{equation}
  \label{rur}
 r' U_{jl}(r',r)r= 
  \frac{2}{\pi}
\!\int_{0}^{\infty} \!\! dk' 
\!\int_{0}^{\infty} \!\! dk\,
  {\cal S}_l(k'r')\,k'\tilde U_{jl}(k',k)k\, {\cal S}_l(kr) ,
\end{equation}
where ${\cal S}_l$ denotes Riccati-Bessel functions given by
${\cal S}_l(x)\!=\!xj_l(x)$.

To summarize the passage of momentum- to coordinate-space
representation of potentials,
starting from known values of the central and spin-orbit terms
in momentum space, 
$\tilde U_{0}(k',k;u)$ and 
$\tilde U_{1}(k',k;u)$, 
we use Eqs. \eqref{m0} and \eqref{m1} to obtain
$\tilde U_{jl}(k',k)$ in Eq.\eqref{Ujl}.
The passage to coordinate space is completed with the double
Fourier transform expressed by Eq.~\eqref{rur}.
The resulting potential is then used to evaluate the
kernel in Eq.~\eqref{ekernel} to obtain scattering waves.
In the above, we denote 
$\tilde U_{0}(k',k;\cos\theta)\!\equiv\!\tilde U_{0}(k',k,\theta)$,
with analogous notation for $\tilde U_1$.

To evaluate the volume integral $J$ of the potential
from its momentum-space representation $\tilde U$ we use
Eq.~\eqref{qtor}, leading to
\begin{equation}
  \label{J}
  J=(2\pi)^3\; \tilde U({\bm k}'\!=\!0,{\bm k}\!=\!0)\;.
\end{equation}
Thus, the volume integral of the potential is 
proportional to its value in momentum space
at ${\bm k'}\!=\!{\bm k}\!=\!0$.
Consistently, in coordinate representation we obtain
\begin{equation}
  \label{Jr}
  J=4\pi\int_{0}^{\infty} r'^2dr'\int_{0}^{\infty}r^2\,dr
  U_{j0}(r',r)\;,
\end{equation}
with $j\!=\!\sfrac{1}{2}$. 
These two forms of $J$ are calculated by the code.

\subsection{Elastic scattering observables}
Here we spell out the formulas used to evaluate the scattering
observables.
Considering collisions of spin-$\sfrac{1}{2}$ nucleons
with spin-\num{0} target, the differential cross section
for unpolarized-beam \textit{NA} scattering is given by
\begin{equation}
\dfrac{d\sigma}{d\Omega} = |g(\theta)|^{2} + |h(\theta)|^{2},
\end{equation}
with the scattering amplitudes $g(\theta)$ and $h(\theta)$  
given by
\begin{subequations}
      \begin{align}
        \label{g}
        g(\theta) &= f_{_C}(\theta) + \frac{i}{4k} 
        \sum_{l=0}^{\infty} \sum_{j} 
        (2j+1) 
        e^{2i\sigma_l}(1-\bar S_{jl}) P_l(\cos\theta)\\
        \label{h}
        h(\theta) &=-\frac{1}{2k} 
        \sum_{l=1}^{\infty} \sum_{j} 
        C_{jl}\,
        e^{2i\sigma_l}(1-\bar S_{jl})\,
        \frac{\partial P_l(\cos\theta)}{\partial\theta}\;.
      \end{align}
\end{subequations}
Summations over $j$ range from 
$|l\!-\!\sfrac{1}{2}|$ to 
$(l\!+\!\sfrac{1}{2})$.
  In the above,
  $\theta$ corresponds to the center-of-momentum deflection
  angle of the projectile, $P_l$ denotes Legendre polynomial, and
  coefficient $C_{jl}$ given by
    \begin{equation}
      C_{jl}=
        \frac{(2j+1)
        \langle\sigma\cdot\ell\rangle_{jl}
        }{2l(l+1)} =
        \left \{
          \begin{array}{rl}
            -1 & \textrm{for $j=l-\sfrac{1}{2}$};\\
            +1 & \textrm{for $j=l+\sfrac{1}{2}$}.\\
          \end{array}
          \right .
    \end{equation}
Additionally, Coulomb amplitude $f_{_C}(\theta)$ is given by
\begin{equation}
  \label{coul}
f_{C}(\theta)=
  \frac{-\eta} {2 k \sin^{2}(\theta/2)}
  \;
  \exp \{ -i\eta\ln[\sin^{2}(\theta/2)] + 2i\sigma_0 \} \;.
\end{equation}
Scattering experiments using polarized beams allow  
measurements of analyzing power $A_y$ and spin rotation 
function $Q$.
These quantities are calculated by  {\small SWANLOP} from
\begin{equation}
 \label{ay}
  A_{y}(\theta) + i\, Q(\theta) =
 \frac{2\, g^{*}(\theta)\,h(\theta) }
      {|g(\theta)|^{2}+|h(\theta)|^{2}}\,.
\end{equation}

Total (integrated) cross sections are evaluated with
\begin{subequations}
  \label{xsections}
\begin{align}    
  \label{SR}
  \sigma_{_R} &= \frac{\pi}{2k^{2}}
        \sum_{l=0}^{\infty} \sum_{j} 
  (2j+1)\left(1-\left|S_{jl}\right|^{2}\right)\;; \\
  \label{SE}
  \sigma_{_E} &= \frac{\pi}{2k^{2}}
        \sum_{l=0}^{\infty} \sum_{j} 
   (2j+1)\left|1-S_{jl}\right|^{2} \;;\\
  \label{ST}
 \sigma_{_T}  &= \frac{\pi}{k^{2}}
        \sum_{l=0}^{\infty} \sum_{j} 
  (2j+1) \left ( 1-\mathrm{Re}\{S_{jl}\}\right ) \;.
\end{align}
\end{subequations}
Here $\sigma_{_R}$, $\sigma_{_E}$ and $\sigma_{_T}$ 
denote reaction, shape-elastic and total cross sections, respectively. 
In these expressions 
$S_{jl}\!=\!\exp[2i(\sigma_l +\bar\delta_{jl})]$.
For proton scattering only the reaction cross section is meaningful,
as both $\sigma_{_E}$ and $\sigma_{_T}$ diverge with increasing 
number of partial waves.

\subsection{General considerations}
Calculations performed by {\small SWANLOP} allow for nucleon
energies of up to \num{1.1}~GeV. Thus, relativistic corrections of
kinematical nature need to be implemented. A brief description
of these corrections are given in \ref{sec:relativity}.
Additionally, proton collisions require the inclusion of 
Coulomb interactions. The model we use is that due to a uniform
charge distribution as described in \ref{sec:coulomb}.
However, the specific subroutine for Coulomb potential evaluation 
can be customized to meet specific requirements.

As guiding rule for the maximum radius of integration, $R_{max}$,
we follow the prescription
\begin{equation}
  \label{rmax}
  R_{max}=r_0A^{1/3} + \bar R\;,
\end{equation}
with $r_0\!=\!1.2$~fm,
$\bar R\!=\!8$~fm, and $A$ the mass number of the target. 
This sets the maximum integration radius about 8~fm further 
away from the surface of the target.
With respect to the maximum orbital angular momentum
to be considered we follow the rule
\begin{equation}
  \label{lmax}
  L_{max} \sim k \,R_{max}\;,
\end{equation}
with $k$ the c.m. momentum.
With the above, collisions of 1~GeV protons off $^{226}$Ra
would lead to $R_{max}\!\approx\!16$~fm, with $L_{max}\!=\!130$.
We stress that these are guiding rules. Actual values for $R_{max}$
and $L_{max}$ may depend on specific features of the 
potential together with the needed precision of observables under study.

Another important consideration is the radial step length $h$ 
to be used by the code to solve the scattering problem.
Here we expect a spatial oscillatory behavior for the wavefunction, 
as driven by the c.m. wavenumber $k$.
In order to keep track of these oscillations we impose 
that each cycle is sampled a certain number of times,
feature which accommodates well to the trapezoidal quadrature
in the radial coordinate.
Keeping control on the dimension matrices to be inverted
together with reasonable accuracy in the calculated observables, 
we have found that half-cycles of the free waves
being sampled by at least six points yields acceptable accuracy. 
With this empirical rule we estimate 
\begin{equation}
  \label{hmax}
  h \lesssim \frac{\pi}{6\,k}\;,
\end{equation}
condition checked by the program which issues a warning message 
if not met.
Thus, for a given $R_{max}$ the value of $h$ is controlled 
by the dimension $N$ of the matrix representing the kernel 
in Eq.~\eqref{kernel}.

An element which also conditions the value of $h$ is the nature 
of the potential. As demonstrated in Ref.~\cite{Arellano2018}, 
microscopic momentum-space potentials when transformed into
coordinate representation exhibit strong oscillating patterns. 
The roughness of these patterns depends on the upper momentum at 
which they are defined in momentum space. 
Interestingly, a reduction (\textit{via} cut off) of the upper 
momenta of the potential yields smoother nonlocal potentials
with the same scattering observables and wavefunctions. 
In the context of Schr\"odinger's wave equation, these smoother 
nonlocal potentials become computationally less demanding in terms 
of the step size $h$.

\section{The package {\small SWANLOP}}
\label{sec:package}

The package is distributed in a single tarred and zipped file named
\texttt{swanlop.tar.gz}.
To unwrap the package apply the command:\\
\indent \texttt{tar -xvfz swanlop.tar.gz} \\
This action will create the directory {\small SWANLOP/}
containing the following file and subdirectories:
\begin{enumerate}
 \item \texttt{./README}\\
   containing instructions to setup the program, 
   prepare inputs and run instructions;
 \item \texttt{./sources/}\\
   subdirectory containing the main program 
    \texttt{swanlop.f}, twenty-six subroutines and 
   twelve functions written in Fortran 90.
   Additionally, it contains a makefile and the executable file;
\item \texttt{./runs/}\\
  subdirectory for inputs, outputs and code execution; and
\item \texttt{./udata/}\\
  subdirectory containing input potentials for testing and reference.
\end{enumerate}

The {\small SWANLOP} package is self-contained, 
independent of any library.
To compile the code, once at subdirectory 
\texttt{./SWANLOP/sources/} 
type
\texttt{make} followed by return key.
This action will create the executable \texttt{swanlop.x} at
\texttt{./SWANLOP/sources/} 

\subsection{Data}
\label{sec:data}
Fundamental constants and unified atomic mass units are stored 
in file \texttt{include\textunderscore\,phys} at subdirectory
\texttt{./SWANLOP/sources/}. Their values are\\

\begin{table}[H]
\begin{center}
\begin{tabular}{ll}
  $\hbar c\! =$\num{197.3269788}~MeV~fm &
  Conversion constant~\cite{PDG2018}\\
  $\alpha\! =$\num{1/137.035999} &
  Fine-structure constant~\cite{PDG2018}\\
  $u\!=$\num{931.494095}  MeV/$c^{2}$ &
  Unified atomic mass unit~\cite{Huang2017} 
\end{tabular}
\end{center}
\end{table}
\noindent
Whenever any of these values is modified delete all 
\texttt{\textasteriskcentered.o} files and re-compile.
Additionally, file \texttt{NucChart} at subdirectory
\texttt{./SWANLOP/runs/} stores mass excess data of 
\num{3436} nuclides,
to obtain masses of the colliding particles during runs.
This data basis has been obtained from 
\textit{The AME2016 atomic mass evaluation} \cite{Huang2017,Wang2017}.

Input files to be prepared by the user to run the code 
are the following:
\begin{center}
\begin{tabular}{ll}
  \texttt{fort.1}: & main input with run specifications;  \\
  \texttt{fort.2}: & (optional) external nonlocal potential; and \\
  \texttt{fort.22}: & (optional) external local potential.
\end{tabular}
\end{center}
\noindent
Follow instructions given at \texttt{SWANLOP/runs/README} to
construct \texttt{fort.1} according to specified requirements.
Further explanations are given in Sec.~\ref{sec:ifiles}.

\subsection{Execution}
\label{sec:exec}
The execution of the program is performed at subdirectory
\texttt{./SWANLOP/runs/}, typing
\begin{center}
\texttt{../sources/swanlop.x} 
\end{center}
followed by return key.
After execution, {\small SWANLOP} generates three outputs by default,
with two additional (optional) outputs if specified. 
These outputs are \texttt{zz.main}, \texttt{zz.xaq}, \texttt{zz.dsdt}, 
\texttt{zz.wave} and \texttt{zz.vrr}, to be described in 
Sec.~\ref{sec:outputs}.

\subsection{Input files}
\label{sec:ifiles}
\subsubsection{Main input}
The main input file is \texttt{fort.1}, consisting of sixteen lines
listed in Table~\ref{tab:fort1}, where we maintain
the notation used in the main code \texttt{swanlop.f}.
For the \texttt{HEADING} entry use any US keyboard character,
excepting empty spaces, slashes (/), 
semicolons (;) and commas (,) as they may trim off any text after 
their occurrence.
The collision is defined with entries \texttt{PROJ}, \texttt{TARGET}
and \texttt{ELAB}, defining the projectile, target and nucleon
beam energy, respectively.
Radial integration specifications are given by \texttt{RMAX} and 
\texttt{NRP}, representing $R_{max}$ and $N$ in Eq.~\eqref{matrixeq}.
The maximum orbital angular momentum $L_{max}$ is specified
by \texttt{LMAX}. 
We refer the reader to \ref{sec:special} for considerations 
on these three entries when potentials are read from file.
\begin{table} [H]
 \begin{center}
 \begin{tabular}{rllll}
   Line & Entry  & Type & Meaning & Values\\
  \hline
  1 & \texttt{HEADING} & Character& Unbroken 70-character job title &\\
  2 & \texttt{PROJ}    & Character& Projectile& $p$ or $n$ \\
  3 & \texttt{TARGET}  & Character& Target specification & e.g. Ca40 \\
   4 & \texttt{ELAB}   & Real& Nucleon beam energy $E_{lab}$& \\
   5 & \texttt{RMAX}   & Real& Maximum integration radius $R_{max}$& \\
  6 & \texttt{NRP}     & Integer& Number of radial points $N$& \\
   7 & \texttt{LMAX}   & Integer& Maximum angular momentum $L_{max}$& \\
  8 & \texttt{ANGMAX,DANG} & Real 
    & Angular array [deg] for $d\sigma\!/\!d\Omega$ 
    & \texttt{ANGMAX}$\leq\!180$ \\
  9 & \texttt{KIN}    & Integer& Relativistic kinematics & 0(no) 1(yes) \\
10 & \texttt{KPOT} &Integer& Potential specification& 0, 1, 2, 3 or 4\\
   11 & \texttt{KADD} & Integer& Addition of local potential& 0(none) 1(read) 2(call)\\
  12 & \texttt{KPRwave} & Integer& Print wavefunctions& 0(no) 1(yes) \\
  13 & \texttt{KPRvrr}  & Integer& Print nonlocal potential& 0(no) 1(yes)\\
   14 & \texttt{DATdsdw} & Character*18& Filename for $d\sigma/d\Omega$ data & \texttt{none} if none\\
   15 & \texttt{DATay} & Character*18& Filename for $A_y$ data & \texttt{none} if none\\
   16 & \texttt{DATqrot} & Character*18& Filename for $Q$ data & \texttt{none} if none\\
  \hline
  \end{tabular}
  \caption{ \label{tab:fort1}
    Entries in \texttt{fort.1} to specify the main task.}
  \end{center}
\end{table}

Parameters \texttt{ANGMAX} and \texttt{DANG} at line 8
specify the angular array for the c.m. angle $\theta$
over which angular scattering observables are to be evaluated. 
Entries are given in degrees, 
with \texttt{ANGMAX} the maximum scattering
angle $\theta$ and \texttt{DANG} the angular step.
If \texttt{ANGMAX}$<\!0$, the program sets the grid internally.
Entry \texttt{KIN} defines the kinematics to be applied in the
\textit{NA} collision. When \texttt{KIN=1}, relativistic 
kinematics is used as described in \ref{sec:relativity}.

Entry \texttt{KPOT} at line 10 
defines the potential to be considered in the run.
The allowed values and meaning are summarized in Table~\ref{tab:kpot}.
We note that under choices \texttt{KPOT=1,2}, the optical potential
is generated internally by the code,
using PB optical model~\cite{Perey1962}
or Tian-Pang-Ma (TPM) parametrization~\cite{Tian2015} of PB model.
 The option to superpose a local potential to nonlocal ones is 
 explained in Sec.~\ref{sec:kadd}.
\begin{table} [H]
\begin{center}
  \begin{tabular} {rl}
    \texttt{KPOT} & Meaning\\
    \hline
    0 & For purely local potential read from file\\
    1 & For Perey-Buck nonlocal model\\
    2 & For TPM parametrization in PB-type model\\
    3 & For coordinate-space nonlocal potential read from file\\
    4 & For momentum-space potential read from file\\
    \hline
  \end{tabular}
  \caption{ \label{tab:kpot}
    Valid options for \texttt{KPOT} and corresponding action.}
\end{center}
\end{table}

Under \texttt{KPOT=0, 3} or 4,
input files \texttt{fort.2} and/or \texttt{fort.22}
containing the potential to be read must be accessible at subdirectory 
\texttt{./SWANLOP/runs/}. 
In Table~\ref{tab:notation} we indicate with checkmarks entries that
must be supplied in the first line of \texttt{fort.2} or \texttt{fort.22}.
Samples of these input files are included in 
subdirectory \texttt{./SWANLOP/udata/.}

  \noindent
   \begin{table} [H]
\begin{center}
    \begin{tabular} {  l l l  c c c }
 \hline
 Entry & Type & Meaning &\multicolumn{3}{c }{\texttt{KPOT}}\\
 \hline
       &      &                                         &0&3&4\\
 \hline
 {\tt ELAB}& real & Nucleon beam energy in MeV
   &\ding{51} 
   &\ding{51}
   &\ding{51}
      \\
 {\tt NAA} & integer & Target mass number
   &\ding{51}
   &\ding{51}
   &\ding{51}
      \\
 {\tt NZZ} & integer & Target proton number
   &\ding{51}
   &\ding{51}
   &\ding{51}
      \\
   {\tt RMAX}& real & Maximum radius in fm                
   &\ding{51}
   &\ding{51}
   &   --   
      \\
 {\tt NRP} & integer & Number of radial points
   &\ding{51}
   &\ding{51}
   &   -- 
      \\
      {\tt LMAX}& integer & Maximum angular momentum
   &   -- 
   &\ding{51}
   &   --
      \\
 {\tt NQF} & integer & Momentum mesh size
   &   --
   &   --
   &\ding{51}
      \\
 {\tt NTH} & integer & Angular mesh size
   &   --    
   &   --    
   &\ding{51}
      \\
 \hline
 \end{tabular}
\caption{ \label{tab:notation}
  Checkmarks on entries that must appear in first line 
  of potential files \texttt{fort.2} and \texttt{fort.22}
  according to \texttt{KPOT} choice.}
\end{center}
      \end{table}

\subsubsection{\texttt{KPOT} option}
\label{sec:kpot}
     Entry \texttt{KPOT} defines the potential to be treated
     by {\small SWANLOP}. There are five possible options
     covering different scenarios.
     We briefly describe actions taken by {\small SWANLOP}
     under each of them.
\begin{enumerate}[label=(\alph*)]
  \item \texttt{KPOT=0}. 
    Option to work with a purely local potential in coordinate space.
    The structure of the potential is assumed as
   \begin{equation}
     \label{vlocal}
     U(r) = U_c(r) +  {\bm\sigma}\!\cdot\!{\bm\ell}\;U_{so}(r)\;.
   \end{equation}
   The terms $U_c(r)$ and $U_{so}(r)$
    are read from file \texttt{fort.22}.
   After the first row 
   the potential must be listed in four columns,
   with an additional (first column) specifying 
   the radial coordinate.
    Accordingly, reading is done as
\begin{verbnobox}
  READ(22,*) ELAB,NAA,NZZ,RMAX,NRP ! First line
Loop_r: DO  K=1,NRP
READ(22,*) r,x0,y0,x1,y1 ! r ReUc ImUc ReUso ImUso
cv0(k) = cmplx(x0,y0)   ! Forms complex Uc
cv1(k) = cmplx(x1,y1)   ! Forms complex Uso
END DO Loop_r
\end{verbnobox}
Here, 
   \texttt{r} denotes the radial coordinate;
   \texttt{x0} denotes $\textrm{Re}\,U_c$;
   \texttt{y0} denotes $\textrm{Im}\,U_c$;
   \texttt{x1} denotes $\textrm{Re}\,U_{so}$; and
   \texttt{y1} denotes $\textrm{Im}\,U_{so}$.
The radial coordinate is given in fm units and the
potential in MeV units.

  \item \texttt{KPOT=1}. 
    Option to apply PB nonlocal model~\cite{Perey1962} with 
    parameters stored internally.
    There is no need to prepare \texttt{fort.2} input file in this case.
    This model has been developed for neutron
    scattering at beam energies between 4 and 24~MeV.
  \item \texttt{KPOT=2}. 
    Option to apply PB-type potential under TPM
    parametrization~\cite{Tian2015}.
    Here also parameters are stored internally,
    without need to prepare \texttt{fort.2} input file.
    This parametrization has been developed for proton
    and neutron scattering,
    at beam energies between 10 and 30~MeV.
  \item \texttt{KPOT=3}. 
    Option to read nonlocal potential in coordinate space,
    $r'U_{jl}(r',r)r$, from file.
    Note that the potential is multiplied by $rr'$.
  Since the potential $U(r',r)$ is expressed in MeV~fm$^{-3}$ units,
  the entry $r'U(r',r)r$ must be given in MeV~fm$^{-1}$ units.
    The potential must be defined over a radial mesh of 
    \texttt{NRP} radial points, evenly spaced, excluding 
    the origin $r\!=\!r'\!=\!0$.
    Since the potential is symmetric under interchange of coordinates,
    $U_{jl}(r',r)\!=\!U_{jl}(r,r')$, information on the full
    matrix can be stored with only its lower triangular part.
    Accordingly, reading proceeds as follows:
\begin{verbnobox}
READ(2,*) ELAB,NAA,NZZ,RMAX,NRP,LMAX ! First line
    LoopL: DO L=0,LMAX
    if(L==0) JA=2       ! Covers J=1/2 only (L=0)
    if(L==1) JA=1       ! Covers J=L-1/2; J=L+1/2 
  LoopJ: DO NS=JA,2
  READ(2,*) LL,AJ       ! Reads L and J
Loop_r1: DO i=1,NRP
Loop_r2: DO j=1,i       ! Lower triangular matrix
READ(2,*) UX,UY         ! Re{U_jl} Im{U_jl}
cvv(i,j) = cmplx(ux,uy) ! Forms complex potential
cvv(j,i) = cmplx(ux,uy) ! Symmetric image
END DO Loop_r2
END DO Loop_r1
  END DO LoopJ
    END DO LoopL
\end{verbnobox}
    After the first line, the potential is listed in
    \texttt{(2*LMAX+1)} triangular blocks,
    each of them preceded by its corresponding $l$ and $j$ 
    (given by {\tt LL} and {\tt AJ}, respectively).

  \item \texttt{KPOT=4}.
 Option to read potential in momentum representation from file. 
 The central component $\tilde U_c(k',k,\theta)$ is 
 stored in the complex matrix \texttt{CPOT0(:,:,:)}, 
 while the spin-orbit component $\tilde U_{so}(k',k,\theta)$ is
stored in the complex matrix \texttt{CPOT1(:,:,:)}.
  These potentials are expressed in MeV~fm$^{3}$ units.
  Angles are expressed in radians and must be listed in ascending order.
Beware of the use of \textsf{implied DO} to read the angular dependence.
Reading in this case proceeds as
\begin{verbnobox}
  READ(2,*) ELAB,NAA,NZZ,NQF,NTH ! First line
  READ(2,*) (AQ(K),K=1,NQF)      ! Momenta [1/fm]
  READ(2,*) (TH(K),K=1,NTH)      ! Angles  [rad]
Loop_k2: DO J=1,NQF
Loop_k1: DO I=1,NQF
READ(2,*) (CPOT0(N,I,J),N=1,NTH) ! U_c(*,i,j)
READ(2,*) (CPOT1(N,I,J),N=1,NTH) ! U_so(*,i,j)
END DO_k1
END DO_k2
\end{verbnobox}
In the above, \texttt{AQ(:)} stores the momentum array 
    (in fm$^{-1}$ units)
over which the potential is defined. 
The elements of this array do not need to be evenly spaced.
Actual calculations of optical potentials in momentum space in 
Ref.~\cite{Arellano2018} follow the rule for the $n$-th element,
$k_n$, given by $k_n\!=\! K_{max}(n/N_Q)^{3/2}$,
with $K_{max}$ below 12~fm$^{-1}$ and $N_Q$ the number of momenta
in the array.
Actually, the value of $K_{max}$ can be diminished significantly 
after the study reported in Ref.~\cite{Arellano2018} on the 
relevance of high momentum components in optical potential models.
With regard to the array \texttt{TH(:)},
this contains the angles $\theta_n$ 
  expressed in radians in the interval $(0,\pi)$
 at which the potential is evaluated.
These elements correspond to those from an \texttt{NTH}-point 
Gaussian quadrature, where its $n$-th element $u_n$ and $\theta_n$
are related through
\begin{equation}
  u_n = \cos\theta_n\;.
\end{equation}
The advantage of this construction is that multipoles of the potential
in momentum space can be obtained without angular interpolation, 
rendering better accuracy to the procedure.
With this, for a given angular array of \texttt{NTH} elements,
the maximum angular momentum to reliably extract multipoles is
\texttt{LMAX=NTH-1}, value used by {\small SWANLOP}.
\end{enumerate}

\subsubsection{Additional local potential}
\label{sec:kadd}
The code offers the possibility to add a local potential
to the one specified by the \texttt{KPOT} option.
This feature is activated when \texttt{KADD=1} or \texttt{KADD=2}
in line 11 of \texttt{fort.1}.
When \texttt{KADD=1} the code reads local potential from file
\texttt{fort.22} with identical format as described
in Sec.~\ref{sec:kpot} for \texttt{KPOT=0},
where the potential is given by its components $U_c(r)$ and $U_{so}(r)$.
The potential must be defined with identical \texttt{RMAX}
and \texttt{NRP} entries, otherwise execution is aborted.
See \ref{sec:optionkadd} for further explanations on this option.
When \texttt{KADD=2} the code calls subroutine 
\texttt{user{\char`_}vloc.f} to evaluate $U_c$ and $U_{so}$.
This subroutine has been coded to be customized by the user.

\texttt{KPRwave} and \texttt{KPRvrr} options are described
in Sec.~\ref{sec:outputs}. 

\subsubsection{Chi-square evaluation}
\label{sec:chi2}
  Entries \texttt{DATdsdw}, \texttt{DATay} and \texttt{DATqrot} 
  denote filenames for experimental measurements of
  $d\sigma\!/\!d\Omega$, $A_y$ and $Q$, respectively.
  These files are formed by three columns, with the first one for
  the c.m. scattering angle, the second for the observable, and the 
  third its error (absolute or percentage).
  Whenever one of these files is declared the code performs
  $\chi^2$ evaluation for the corresponding observable,
  recording results in the main output.
  If no $\chi^2$ evaluation is to be performed in any of these
  observables, then \texttt{none} has to be specified in the 
  corresponding entry.

\subsection{Output files}
\label{sec:outputs}

\begin{enumerate}[label=(\alph*)]
  \item {\tt zz.main} :\hspace{3pt}
     Main output of the code recording collision specifications,
     volume integral per nucleon of local and nonlocal potentials,
     phase-shifts, total cross sections and angular scattering
     observables.

   \item {\tt zz.xaq} :\hspace{3pt}
 Plot-ready output composed of seven columns recording:
    1) Center-of-momentum scattering angle $\theta$;
    2) Momentum transfer $q\!=\!2k\sin(\theta/2)$ in fm$^{-1}$ units;
    3) Momentum transfer $q$ in MeV/$c$ units;
    4) Differential cross section $d\sigma\!/\!d\Omega$ in mb/sr units;
    5) Analyzing power $A_y$;
    6) Spin rotation function $Q$; and
    7) Ratio-to-Rutherford differential cross section.

   \item {\tt zz.dsdt} :\hspace{3pt}
 Plot-ready output composed of four columns recording:
    1) Center-of-momentum scattering angle $\theta$;
    2) Mandelstam $-t$ invariant in (GeV/c)$^2$ units
       ($t\!=\!-q^2$); 
    3) Differential cross section $-d\sigma\!/\!dt$ 
       in mb~GeV$^2$/$c^2$ units; and
    4) Ratio-to-Rutherford differential cross section.
This is a common convention adopted in high-energy scattering 
    experiments~\cite{Korolev2018}.

  \item {\tt zz.waves} :\hspace{3pt}
    \textit{Optional}
    output containing scattering waves $u_{jl}(r)$ as functions
    of the radial coordinate $r$. 
    This output is generated when \texttt{KPRwave=1},
    in line 12 of \texttt{fort.1}.
    Partial waves are listed in {\tt LMAX+1} consecutive blocks,
    each of them defining the orbital angular momentum $l$ and
    number of radial points {\tt NRP}. 
    The block is completed with {\tt NRP} lines, in seven columns
    as follows

    \noindent
    \begin{tabular}{ccccccc}
      l &
      r &
      Re  $u_{-}$   &
      Im  $u_{-}$   &
      Re  $u_{+}$   &
      Im  $u_{+}$   &
      $\textstyle{\frac{1}{k}}F_{l}(\eta,r)$ \\
      $\vdots$ &
      $\vdots$ &
      $\vdots$ &
      $\vdots$ &
      $\vdots$ &
      $\vdots$ &
      $\vdots$ 
    \end{tabular}\\
Subscripts $\pm$ in $u$ denote $j\!=\!l\!\pm\!\sfrac{1}{2}$.
First and second columns correspond to
orbital angular momentum and
radial coordinate $r$ in fm units, respectively;
third and fourth columns correspond to
    $\textrm{Re }\!\{u_{jl}\}$ and
    $\textrm{Im }\!\{u_{jl}\}$
    ($j\!=\!l\!-\!\sfrac{1}{2}$),
    respectively; 
fifth and sixth columns correspond to
    $\textrm{Re }\!\{u_{jl}\}$ and
    $\textrm{Im }\!\{u_{jl}\}$
    ($j\!=\!l\!+\!\sfrac{1}{2}$),
    respectively; and
seventh column corresponds to the undistorted
    Coulomb wave $F_{l}(\eta,r)/k$ in Eq.~\eqref{ekernel}.
    All waves are given in fm units.

  \item {\tt zz.vrr} :\hspace{3pt}
    \textit{Optional} output containing the nonlocal potential 
    $rr'U_{jl}(r',r)$ as function of the radial 
    coordinates $r$ and $r'$.
    This file is generated under \texttt{KPRpot=1}, 
    in line 13 of \texttt{fort.1}.
    The structure of this output file for $r'U_{jl}(r',r)r$
    is identical to the one described in Sec.~\ref{sec:ifiles},
    under option {\tt KPOT=3}.
    Note also that the potential is being multiplied by $rr'$.

    \end{enumerate}

\subsection{Credits}
Two subroutines in {\small SWANLOP} package have been developed 
by other authors. 
The first one, \texttt{coulfg.f}, has been
developed by A.~R.~Barnett to calculate 
regular and irregular Coulomb functions~\cite{Barnett1982}. 
The second one, \texttt{seval\textunderscore\,c.f}, 
corresponds to an adaptation of the cubic spline interpolation 
routine by Moreau~\cite{Moreau2020},
based on Ref.~\cite{Forsythe1977} by Forsythe.

\section{Benchmarks}
\label{sec:benchmark}

In this section we study the accuracy of {\small SWANLOP},
illustrate its convergence features and 
present comparison with {\small SIDES} package~\cite{Blanchon2020}.
As stated in Eq.~\eqref{matrixeq},
after the construction of the kernel over a uniform grid of
$N$ radial points,
the scattering problem is reduced to a matrix equation for the
wavefunction.
For the construction of the kernel we use trapezoidal quadrature,
conveying an estimated error 
$\sim\! R^3 f''(r_m)/N^2$,
with $R$ the maximum radial coordinate, $N$ the number of 
points involved, and $r_m$ a radial coordinate within the range
at which $f''$ is extreme.
Here $f$ is any of the integrands in Eq.~\eqref{kernel}, 
either as function of $r'$ or $r''$.
We now examine how this trend gets manifested in actual applications.
In what follows we denote the radial step size ($dr$) by $h$.

\subsection{Comparison against separable analytic solution}
\label{sec:exact}
Separable potentials offer the possibility of providing with
analytic solutions in closed forms
for the scattering matrix and implied scattering observables.
In this section we assess the ability of {\small SWANLOP}
to reproduce such closed-form results 
with focus on $s$-wave total cross sections. 

Following Ref.~\cite{Bagchi1974}, let us consider
the rank-1 separable potential  
$U(r',r)\! =\! \lambda \,\xi_n(r)\xi_n(r')$,
with form factor defined as
\begin{equation}
  \label{fr}
  \xi_n(r) = (\alpha r)^n \frac{e^{-\alpha r}}{r}\;.
\end{equation}
Here $\lambda$ is given in units of MeV~fm$^{-1}$.
In \ref{sec:analytic} we provide closed-form expressions for
the $S$ matrix in the case of form factors as in Eq.~\eqref{fr},
for the cases $n\!=\!1$ and 2.
We apply these results considering the targets
$^{16}$O, $^{72}$Ge and $^{198}$Hg, with nuclear radii $R_{_A}$ of
3, 5, and 7~fm, respectively.
Their respective strength $\lambda$ are calibrated to give
volume integral of the potential per nucleon 
$J/A\!=\!-400$~MeV~fm$^{3}$. 
The resulting values for $\alpha$ and $\lambda$
obtained from Eqs.~\eqref{joa} and \eqref{rms} are summarized 
in Table~\ref{tab:Tbagchi}.\\
\begin{table}[ht]
\begin{center}
\begin{tabular}{c| c c c }
  \hline
 Target & n &  $\alpha$ [ fm$^{-1}$] & $\lambda$ [ MeV fm$^{-1}$ ] \\
\hline
\hline
  \multirow{2}{*}{$^{16}$O}
  &
  1 & \num{1.4907} &  \num{-50.00}  \\
  &
  2 & \num{1.9245} &  \num{-15.44} \\
\hline
  \multirow{2}{*}{$^{72}$Ge}
  &
  1 & \num{0.8944} &  \num{-29.18}  \\
  &
  2 & \num{1.1547} &  \num{ -9.01} \\
\hline
  \multirow{2}{*}{$^{198}$Hg}
  &
  1 & \num{0.6389} &  \num{-20.89}  \\
  &
  2 & \num{0.8248} &  \num{ -6.45} \\
\hline
\end{tabular}
  \caption{ \label{tab:Tbagchi}
    Parameters used for analytic solutions}
\end{center}
\end{table}
These values were applied in {\small SWANLOP}
for neutron-nucleus elastic scattering at energies ranging from 5 up to 
\num{1100}~MeV. 
This fictitious scenario is conceived with the sole purpose
to test the accuracy of the code over a wide range of energies.
In order to allow for interference between real and imaginary
components, the strengths used in these tests are made complex 
through $\lambda\!\to\!(1+i/4)\lambda$.

\begin{center}
  \begin{figure}[ht]
  \includegraphics[angle=-90,width=\linewidth]{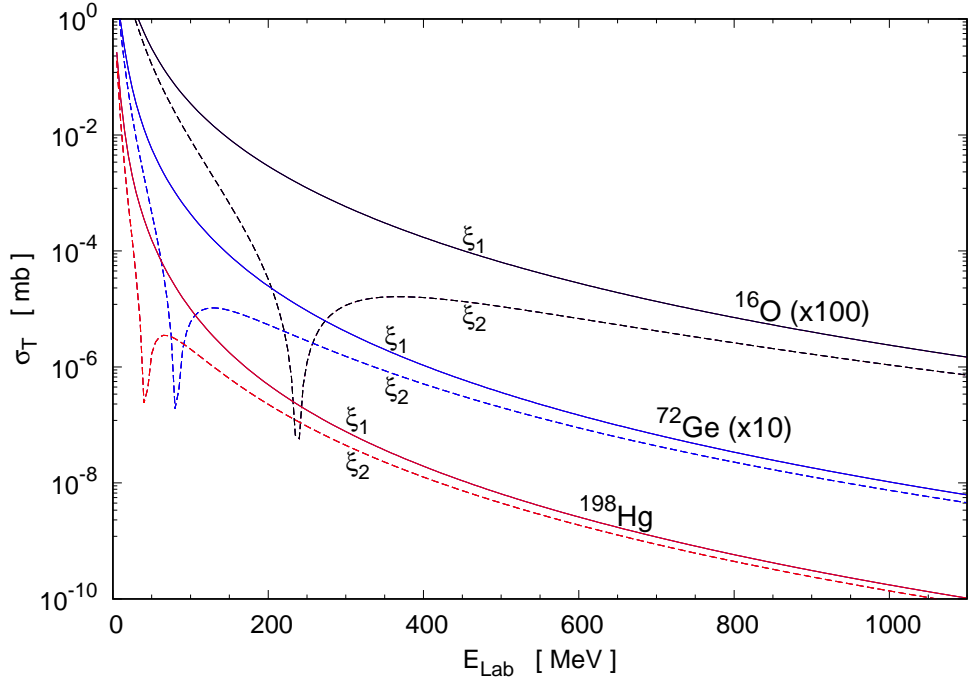}
 \caption{ \label{fig:stotal}
   $s$-wave total cross section as function of laboratory
    energy $E_{Lab}$ for neutron-nucleus scattering from
    $^{16}$O (black curves),
    $^{72}$Ge (blue curves), and
    $^{198}$Hg (red curves).
    For each target and form factor, 
    plots include analytic results together with
    {\small SWANLOP} results using $h\!=\!0.050$, 
    \num{0.075} and \num{0.100}~fm.
    No visual distinction is observed on each case.
    }
  \end{figure}
\end{center}

In Fig.~\ref{fig:stotal} we present results for the $s$-waves 
total cross section $\sigma_{_T}$ based on the numerical solution 
provided by {\small SWANLOP} and the analytic results 
expressed by Eqs.~\eqref{xik}, \eqref{xk} and \eqref{smatrix}. 
Relativistic kinematics has been used throughout.
For clarity, curves associates to $^{16}$O and $^{72}$Ge have 
been up-shifted by factors of 100 and 10, respectively.
Curves labeled with $\xi_1$ (solid) and 
$\xi_2$ (dashed) indicate the form factor used.
The step length $h$ used by {\small SWANLOP} 
in these applications are \num{0.050}, \num{0.075} 
and \num{0.100}~fm.
As observed all curves for $\sigma_{_T}$,
for a given target and form factor $\xi$,
become indistinguishable to the eye,
with $\xi_1$ leading to a monotonic descent.
Results based on $\xi_2$ exhibit sharp minima at
$E_{Lab}$ near 40, 80 and 240~MeV.

\begin{center}
  \begin{figure}[ht]
  \includegraphics[angle=-90,width=\linewidth]{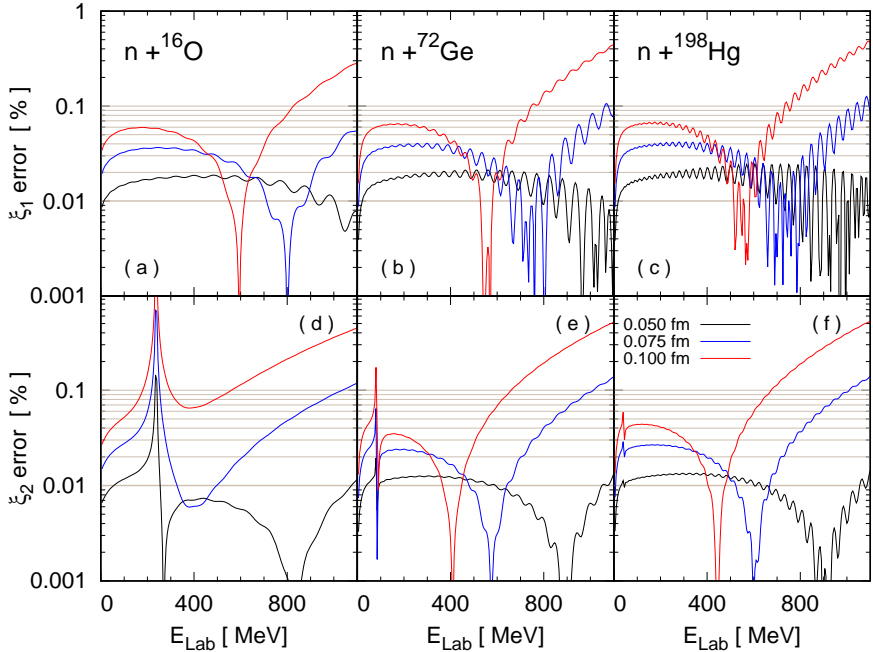}
 \caption{ \label{fig:errors}
   Percentage error for $s$-wave total cross section as function 
   of $E_{Lab}$ for results in Fig.~\ref{fig:stotal}.
   Panels (a), (b) and (c) correspond to $\xi_1$, whereas
    panels (d), (e) and (f) correspond to $\xi_2$.
   Black, blue and red curves denote radial step length of
   \num{0.050}, \num{0.075} and \num{0.100}~fm, respectively.
    }
  \end{figure}
\end{center}

In Fig.~\ref{fig:errors} we present the percentage error
of the numerical solutions obtained with {\small SWANLOP}
relative to the analytic ones.
Panels (a), (b) and (c) show comparisons under $\xi_1$
for $^{16}$O, $^{72}$Ge, and $^{198}$Hg, respectively.
Analogously, panels (d), (e) and (f) show comparisons under $\xi_2$,
for the respective targets.
Black, blue and red curves denote radial step length of
\num{0.050}, \num{0.075} and \num{0.100}~fm, respectively.

Overall, we note that the errors of results from {\small SWANLOP} 
differ from the analytic solution by around \num{0.02}\%,
except for the case $^{198}$Hg under $\xi_1$, where the error
is slightly higher ($\sim\!0.03$~\%).
This overall trend is also broken in the case of form factor
$\xi_2$ in the vicinity of the sharp minima observed
in Fig.~\ref{fig:stotal}. 
Away from these minima,
after observing the errors of solutions based 
on $h\!=\!0.100$~fm (red curves)
we notice that the accuracy of the numerical solution 
remains better than \num{0.1}~\% up to energies
nearing \num{600} to \num{800}~MeV. 
Beyond these energies the accuracy deteriorates monotonically
up to about 0.5\%.
We note that the neutron wavenumber for $E_{Lab}\!=\!700$~MeV
is about \num{6.6}~fm$^{-1}$.
Above this energy the product between the wavenumber and radial
spacing $h$ yields $kh\!\gtrsim\! 0.66$, above the border
of criterion set by Eq.~\eqref{hmax} for $h$.

\begin{center}
  \begin{figure}[ht]
  \includegraphics[angle=-90,width=\linewidth]{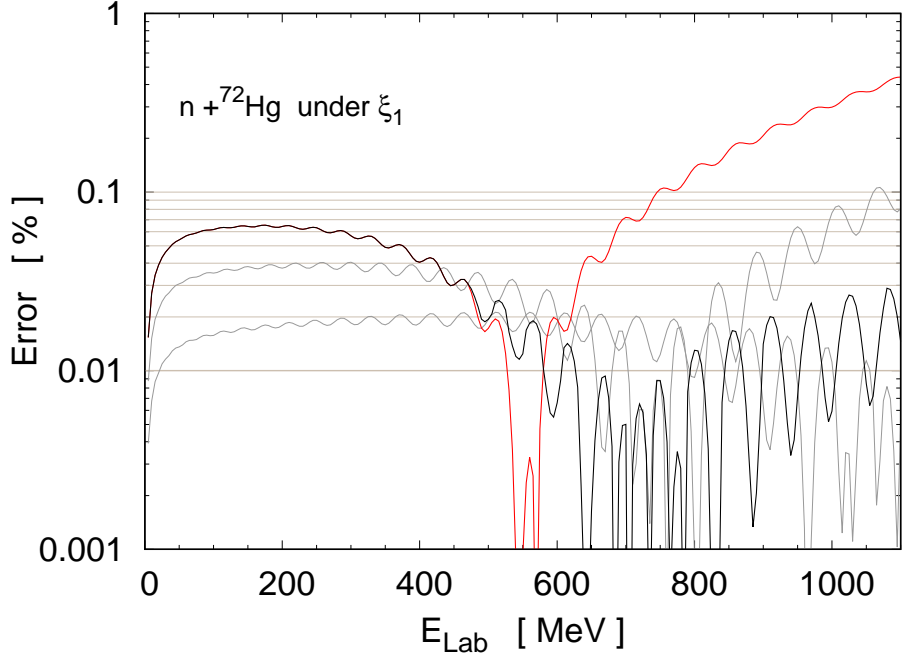}
 \caption{ \label{fig:pi6}
   Percentage error relative to analytic results
   for $s$-wave total cross section,
   as functions of $E_{Lab}$,
   for $^{72}$Hg($n,n$) scattering under separable form factor $\xi_1$. 
   Red curve uses $h\!=\!0.1$~fm, whereas black curve diminishes
   $h$ according to Eq.~\eqref{hmax}.
    }
  \end{figure}
\end{center}

We have analyzed the implications of the above criterion for 
$kh$ in the particular case of $^{72}$Ge under form factor $\xi_1$.
In the analysis we start with $h\!\equiv\!h_0\!=\!0.1$~fm 
at the lowest energy.
As the energy increases we check the value of the product
$\delta\varphi\!=\!kh$, which also increases.
When $\delta\varphi\!>\!\pi/6$, the step length is reset to
$h\!=\!\pi/6k$, keeping $R_{max}$.
Results from this analysis are summarized in Fig.~\ref{fig:pi6},
where we plot percent errors as functions of $E_{Lab}$.
Red curve is based on $h_0$, whereas
black curve is based on $h\!=\min\{h_0,\pi/6\,k\}$.
The two pale curves, 
corresponding to step length of \num{0.075} and \num{0.050}~fm,
are included for reference.
From the actual outputs we find that the departure of the black 
from the red curve takes place at $E_{Lab}\!=\!470$~MeV, 
energy at which $\delta\varphi$ surpasses $\pi/6$. 
Beyond this energy the use of $h\!=\!\pi/6\,k$,
results in errors below \num{0.03}\% ending up in between 
the two pale curves at 1.1~GeV,
where the corrected step length is $h\!\approx\!0.06$~fm.

The preceding analysis has to be taken as informative.
This is so mainly because a rank-1 separable nonlocal model 
is an oversimplification of realistic ones.
As a matter of fact, all applications made in this 
sub-section take $R_{max}\!=\!4 R_{_A}$.
In the case of $^{198}$Hg this means $R_{max}\!=\!28$~fm,
well above the 17~fm prescribed by Eq.~\eqref{rmax}.
The reason in doing so was the imperative need to identify 
the conditions under which {\small SWANLOP} results get 
reasonably close to the analytic results.
For realistic applications, however, 
the prescription given by Eq.~\eqref{rmax} remains adequate.
Beyond these remarks, we have shown that {\small SWANLOP} results,
representing numerical solutions for exact scattering waves 
in the context of Schr\"odinger equation, 
agree with analytic solutions within \num{0.02}\%, using
$h\!=\!0.05$~fm.
An improvement beyond these estimates goes beyond the scope
of this work.

\subsection{Convergence under step size for nonlocal
optical-model potentials}
In this section we illustrate convergence features of the code
as a function of the step length $h$ of the solutions,
considering PB-type nonlocal optical model
as well as momentum-space potentials obtained from microscopic 
calculations.
In these applications we focus on differential observables
for \textit{pA} elastic scattering.
The number of partial waves to consider follow the rule given
by Eq.~\eqref{lmax}.

\subsubsection{TPM nonlocal model for \textit{pA} scattering at
\num{30.3}~MeV}
\begin{center}
  \begin{figure}[ht]
  \includegraphics[angle=-90,width=\linewidth]{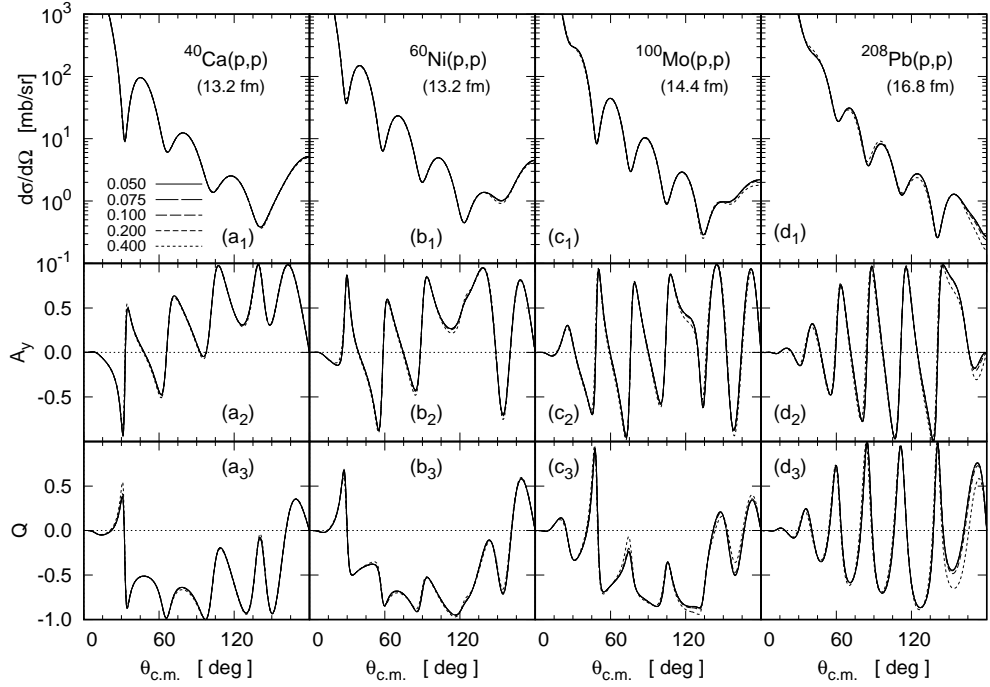}
 \caption{ \label{fig:TPM30}
    Results from {\small SWANLOP} for TPM nonlocal model
    applied to \num{30.3}-MeV proton scattering off $^{40}$Ca (a), 
    $^{60}$Ni (b), $^{100}$Mo (c), and $^{208}$Pb (d).
    Legend labels in panels ($a_1$) denote $h$ in fm units.
    }
  \end{figure}
\end{center}

We now make use of {\small SWANLOP} to study proton scattering
at 30.3~MeV using TMP parametrization of PB nonlocal model.
The selected targets are 
$^{40}$Ca, $^{60}$Ni, $^{100}$Mo, and $^{208}$Pb.
\label{sec:TPM30MeV}
In Fig.~\ref{fig:TPM30} we plot $d\sigma\!/\!d\Omega$, $A_y$ and $Q$
as functions of the c.m. scattering angle for proton scattering off
$^{40}$Ca (a),
$^{60}$Ni (b),
$^{100}$Mo (c), and
$^{208}$Pb (d).
The values used for $R_{max}$ on each case are indicated in parenthesis,
chosen to match step sizes of \num{0.050}, \num{0.075}, 
\num{0.100}, \num{0.200} and \num{0.400}~fm.
Legend labels in frame (a$_1$) indicate the radial step in fm units.
These figures illustrate stable convergence of the results as
the step size diminishes, involving medium-size and large targets.
Actually, only those cases with $h\!=\!0.2$ and 
\num{0.4}~fm depart slightly
from the rest, indicating that $h\!=\!0.1$~fm is safe enough 
for {\small SWANLOP} to obtain reliable observables under TPM 
nonlocal model.

To supplement these findings, in Table~\ref{tab:sigmaR} we tabulate
the calculated reaction cross sections for \textit{pA} scattering
at 30.3~MeV, for the same targets and values of $h$ 
included in Fig.~\ref{fig:TPM30}.
The first column represents the integration step length. 
We note that targets $^{60}$Ni, $^{100}$Mo and $^{208}$Pb
exhibit no variation in $\sigma_{_R}$ for $h\!\leq\!0.1$~fm.
The case of $^{40}$Ca exhibits variations in the fourth significant
figure, of the order of \num{0.02}\%, comparable to errors
relative to the analytic solutions discussed in
Sec.~\ref{sec:exact}.
\begin{table}[H]
\begin{center}
  \captionsetup{width=0.5\textwidth}
    \begin{tabular}{c|cccc}
    \hline
      $h$ [fm]    &    \multicolumn{4}{c}{$\sigma_{_R}$ [b]}\\
    \hline
       &$^{40}$Ca &  $^{60}$Ni &  $^{100}$Mo &  $^{208}$Pb\\
    \hline
    \num{0.050} & \num{0.9162} &\num{1.075} &\num{1.336}&\num{1.589}\\
    \num{0.075} & \num{0.9163} &\num{1.075} &\num{1.336}&\num{1.589}\\
    \num{0.100} & \num{0.9164} &\num{1.075} &\num{1.336}&\num{1.589}\\
    \num{0.200} & \num{0.9174} &\num{1.076} &\num{1.337}&\num{1.590}\\
    \num{0.400} & \num{0.9214} &\num{1.080} &\num{1.341}&\num{1.593}\\
    \hline
    \end{tabular}
\caption{\label{tab:sigmaR}
 Calculated reaction cross sections $\sigma_{_R}$ for 
  \textit{pA} scattering at \num{30.3}~MeV as functions 
  of the step length $h$.  TPM parametrization is used.}
\end{center}
\end{table}

\subsubsection{Momentum-space potential for nucleon scattering 
off $^{40}$Ca at 80 MeV}
\begin{center}
  \begin{figure}[ht]
    \begin{center}
  \includegraphics[angle=-90,width=\linewidth]{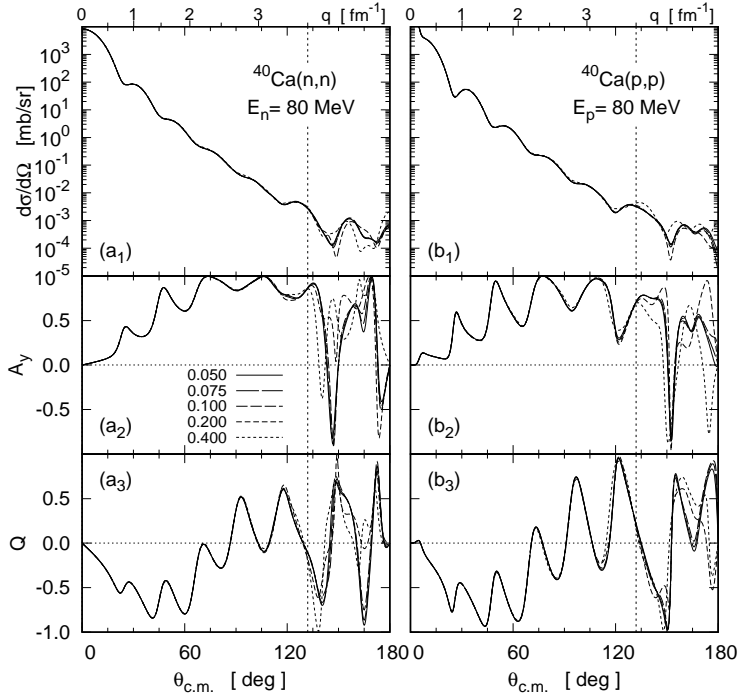}
    \end{center}
 \caption{ \label{fig:40Ca80MeV}
    Scattering observables obtained by {\small SWANLOP}
    for 80-MeV proton and neutron collisions off $^{40}$Ca.
    Microscopic nonlocal potential obtained in momentum 
    space within $g$-matrix folding model of Ref.~\cite{Aguayo2008}.
    Legend labels refer to $h$ in fm units.
    }
  \end{figure}
\end{center}

Along the same line as in the preceding section,
we now consider neutron and proton scattering off $^{40}$Ca at 80 MeV.
In this case the potential is defined in momentum space,
evaluated at 28 angles generated from Gaussian quadrature.
Radial integration is up to $R_{max}\!=\!13.2$~fm,
under \texttt{KPOT=5}.
Scattering calculations by {\small SWANLOP} were performed 
considering $h\!=\!0.050$, \num{0.075}, \num{0.100}, \num{0.200}
and \num{0.400}~fm. 
In Fig.~\ref{fig:40Ca80MeV} we plot results for 
$d\sigma\!/\!d\Omega$ ($a_1,b_1$), 
$A_y$ ($a_2,b_2$) and $Q$ ($a_3,b_3$) as functions
of the scattering angle $\theta_{c.m.}$.
The upper scale denotes momentum transfer $q$,
with the vertical dotted line at $q\!=\!3.5$~fm$^{-1}$ 
drawn for reference.
As in the \num{30.3}~MeV applications, 
\textit{NA} scattering observables calculated with $h\!\leq\!0.1$~fm 
become difficult to distinguish from one another, 
from which we infer that $h\!=\!0.1$~fm enables converged results.
For $h\!>\!0.1$~fm, instead, observables at $q\!>\!3.5$~fm$^{-1}$
depart from the rest as
dotted and short-dashed curves become distinguishable. 
Momentum transfers of about 4~fm$^{-1}$ is a typical upper 
limit of scrutiny for \textit{NA} scattering at intermediate 
energies~\cite{Ray1992}, 
i.e. nucleon beam energies from a few hundred MeV to about \num{1}~GeV.

\begin{center}
  \begin{figure}[ht]
    \begin{center}
      \includegraphics[angle=-90,width=\linewidth]{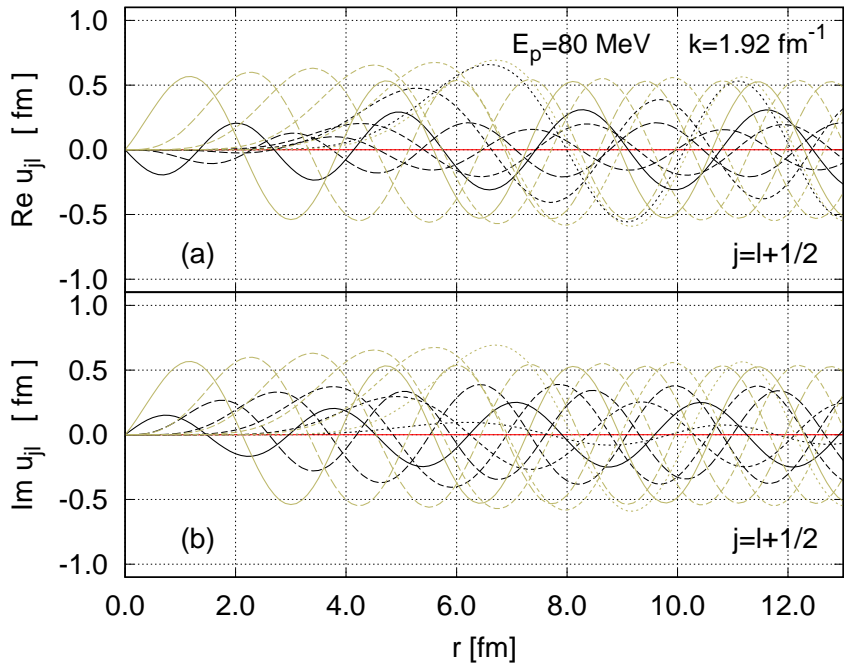}
    \end{center}
 \caption{ \label{fig:waves}
    Scattering waves (black curves) off nonlocal potential obtained 
    by {\small SWANLOP} for $^{40}$Ca($p,p$) at 80~MeV.
    Panels (a) and (b) show the real and imaginary component
    of $u_{jl}$, respectively.
    The potential (originally in momentum-space) 
    corresponds to the same as used in Fig.~\ref{fig:40Ca80MeV}.
    Colored curves correspond to free Coulomb waves.
    Plots include even-number orbital angular momentum, 
    with $l\!\leq\!10$.
    }
  \end{figure}
\end{center}

To complete this application at 80~MeV, in Fig.~\ref{fig:waves}
we plot the scattering waves for $^{40}$Ca($p,p$), based
on the same nonlocal potential calculated in momentum space.
The beam momentum in this case is $k\!=\!1.92$~fm$^{-1}$, 
and select stretches states $j\!=\!l+\sfrac{1}{2}$,
with $l\!\leq\!10$. 
In panel (a) we show the real component of $u_{jl}$ whereas
in panel (b) we plot its imaginary component.
In these plots we consider waves with even $l$, 
with $s$ waves plotted with solid lines. 
Waves with $l\!\geq\!2$ are plotted with segmented curves,
with decreasing dash-length as $l$ increases.
Colored curves represent undistorted incoming waves
$F_l(kr)/k$ included here as reference in both panels.
With this figure we intend to highlight the capability
of {\small SWANLOP} to calculate scattering waves 
in collision described by momentum-space potentials,
being this the first open code in doing so.

\subsection{Comparison with {\small SIDES}}
\label{sec:sides}
We now proceed to compare results for scattering observables obtained
from {\small SWANLOP} and {\small SIDES}. 
As mentioned earlier, {\small SIDES} is a package developed
to solve Schr\"odinger integro-differential equation in the
presence of nonlocal potentials using finite
differences techniques~\cite{Blanchon2020}.
In the applications we pursue here we consider \textit{pA} 
scattering with proton energies of 200~MeV, 700~MeV and 1~GeV.
The targets to consider are $^{12}$C, $^{40}$Ca, $^{90}$Zr and $^{208}$Pb.

The nonlocal optical potentials for these processes 
are obtained from momentum-space calculations following 
Refs.~\cite{Aguayo2008,Arellano2002}.
Specifically, applications at \num{200}~MeV are based on 
density-dependent
$g$-matrix folding model, with full account of the genuine $g$
matrix off shell.
At \num{0.7} and 1~GeV we use the off-shell $t\rho$ approximation.
Relativistic kinematics in the calculation of the potential is 
included together with the account for hadronic absorption in the bare
\textit{NN} interaction above pion-production 
threshold~\cite{Arellano2002}.
The nonlocal one-body mixed densities are obtained within the Slater
approximation~\cite{Arellano1990} from local neutron and proton 
densities of the targets.
These radial densities are obtained from self-consistent
Hartree-Fock-Bogoliubov calculations with the
Gogny force~\cite{Decharge1980}.
Once the momentum-space potential is calculated, 
{\small SWANLOP} generates its coordinate representation to be used
by {\small SIDES}.

\begin{center}
  \begin{figure}[ht]
  \includegraphics[angle=-90,width=\linewidth]{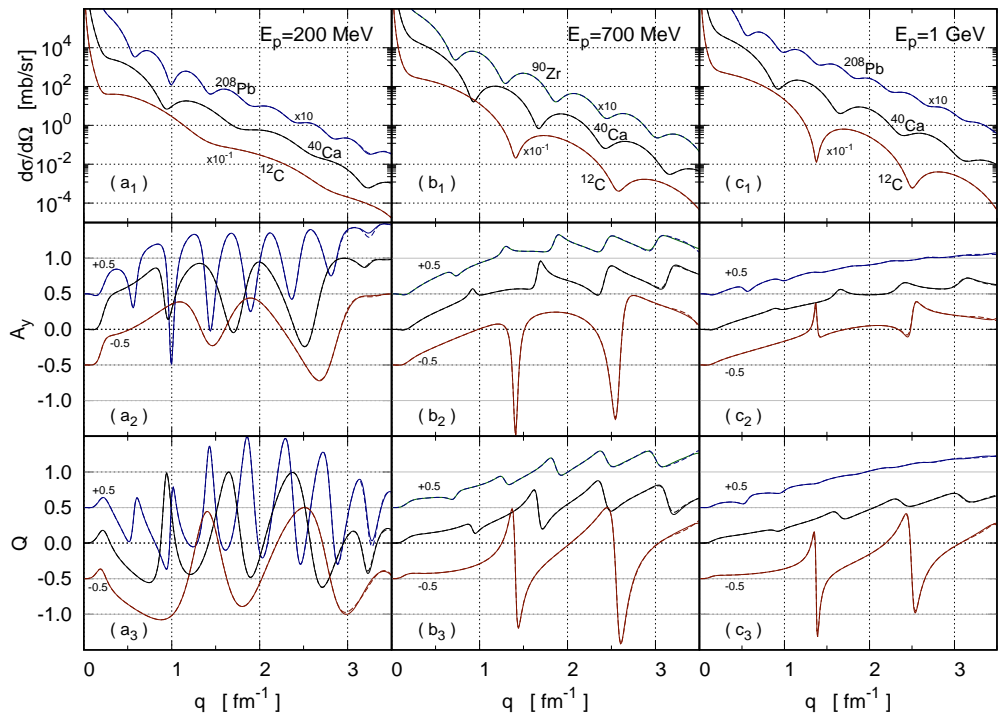}
 \caption{ \label{fig:vsSIDES}
    Scattering observables as functions of the momentum transfer
    $q$ obtained by
    {\small SWANLOP} (solid curves) and 
    {\small SIDES} (dashed curves)
    from microscopic nonlocal potentials.
    Proton elastic scattering at \num{200}~MeV, \num{700}~MeV and 1~GeV.
    See main text for explanation of each frame and information
    on the potentials being used.
    }
  \end{figure}
\end{center}

In Fig.~\ref{fig:vsSIDES} we plot 
---as functions of the momentum transfer $q$---
the differential cross section $d\sigma\!/\!d\Omega$ (upper row),
analyzing power $A_y$ (middle row), and
spin rotation function $Q$ (lower row),
obtained from 
{\small SWANLOP} (solid curves) and
{\small SIDES} (dashed curves).
Columns (a), (b) and (c) correspond to proton energies $E_p$
of \num{200}~MeV, \num{700}~MeV and 1~GeV, respectively.
To avoid superposition of curves in frames ($a_1$), ($b_1$) and ($c_1$), 
results for $d\sigma\!/\!d\Omega$ in the cases of
$^{208}$Pb and $^{90}$Zr have been multiplied by $10$,
whereas those for $^{12}$C have been multiplied by $10^{-1}$.
Similarly, $A_y$ 
for $^{208}$Pb and $^{90}$Zr have been up-shifted by \num{+0.5},
while those
for $^{12}$C are down-shifted by the same amount (\num{-0.5}).
Identical considerations are made for $Q$ in the lower row.

As observed, the agreement between {\small SWANLOP} and
{\small SIDES} results is quite satisfactory,
where in most cases the curves from the two packages overlap
each other.
Some slight differences are observed for $Q$ at 200~MeV
in the case of $^{12}$C($p,p$) in panel ($a_3$), around the 
minimum at $q\!\approx\!3$~fm$^{-1}$.
This is despite the radical difference in the methods applied 
by the two packages, 
with {\small SIDES} using finite difference
techniques to solve the integro-differential equation, 
while {\small SWANLOP} inverts $(1\!-\!\mathbb{K})$ 
in Eq.~\eqref{matrixeq} to obtain the scattering wavefunction.

\subsection{Performance}
\label{timing}
The actual time of execution of the code will depend upon
the speed of the machine under use, in addition to the
potential to be considered. 
However, we have 
found that the CPU run time $\tau$ with maximum 
angular momentum $L_{max}$, using $N$ radial points can
be estimated with
\begin{equation}
  \label{time}
  \tau = \tau_0\,(2L_{max}+1)\,N^3\;.
\end{equation}
The base time $\tau_0$ depends on the machine.
For a 2.6~GHz {Intel\textregistered} {Core\texttrademark} i7
processor used for all
{\small SWANLOP} applications included in this work,
we obtain $\tau_0\!=\!\num{11.5}$~ns.
This is considering potentials in coordinate or momentum space read
from file, suppressing print out of waves and potentials.
With this, the run time for $^{208}$Pb($p,p$) at 1~GeV 
using $N\!=\!320$, and $L_{max}\!=\!129$, would take $\tau\!=\!98$~s,
while the actual run time is \num{97}~s.
In the case of $^{12}$C at \num{200}~MeV with $N\!=\!110$, 
and $L_{max}\!=31$, Eq.~\eqref{time} yields $\tau\!=\!1$~s,
whereas the actual run time is 1.5~s.
All TPM applications in Fig.~\ref{fig:TPM30} with $h\!=\!0.1$~fm
take between 1 and 2~s. 
For PB-type potentials calculated internally, Eq.~\eqref{time} 
for $\tau$ becomes inadequate above 50~MeV beam energy
due to preponderance of time dedicated to multipole calculations.

We note that the total CPU run time $\tau$ depends 
on $L_{max}$ and $N$, both quantities being
guided by Eqs.~\eqref{rmax}, \eqref{lmax} and \eqref{hmax}.
In order to keep the focus on broad energy applications, 
with most targets in the nuclear chart,
covering all scattering angles conditioned by maximum momentum 
transfer $q\!\sim\!4$~fm$^{-1}$,
we have made no effort to optimize these quantities.
Specific uses of the code, however, may allow to relax some of 
these parameters without compromising precision in observables of
interest.

\section{Summary and conclusions}
\label{sec:summary}
We have introduced the self-contained package 
{\small SWANLOP} 
aimed to obtain accurate solutions for \textit{NA} elastic scattering 
under nonlocal potentials 
for spin-zero target nuclei.
The solution is theoretically motivated by Ref.~\cite{Arellano2019},
where scattering waves are obtained from 
a Lippmann-Schwinger type
integral equation for the scattering waves.
Its numerical implementation involves finite matrices over a spatial 
mesh, obtaining scattering waves by direct matrix inversion.
Scattering observables such as differential cross sections 
$d\sigma\!/\!d\Omega$, analyzing power $A_y$ and spin rotation function
$Q$, in addition to integrated cross sections are calculated.
The code offers the possibility to treat local and nonlocal potentials,
or admixture of both. 
Additionally, the code is capable of handling potentials in 
momentum space. 
This is an important feature since developments of microscopic 
or \textit{ab-initio} models tend to evolve independently from 
different groups, mainly due to differences in the representation
of their \textit{NA} interactions.
With the code {\small SWANLOP} it becomes possible, at least,
to study scattering waves from those momentum-space potentials
and compare them with those obtained in coordinate space.

Benchmark studies were carried out at nucleon energies from
few MeV up to \num{1.1}~GeV, 
including light-, medium-mass and heavy targets,
leading to consistent and reliable results. 
These tests also include comparison of
results obtained from the code with those from analytic closed-form
expression, where accuracy within \num{0.02}\% is obtained.
We have also performed comparisons of angular scattering observables
obtained from the package {\small SIDES}~\cite{Blanchon2020}, 
at proton beam energies of \num{200}, \num{700} and \num{1000}~MeV, 
for light and heavy targets.
Results from these applications show remarkable consistency 
between these two packages.

The calculation of scattering waves in \textit{NA} collisions
in the context of nonlocal potentials, 
superposed to the long-range Coulomb interaction, 
has been longstanding problem where specific solutions have been 
introduced under different assumptions on the nature of the nonlocality.
These assumptions are either made explicit by their authors
or made implicit in the adopted calculational scheme.
In the case of momentum-space optical potentials, 
codes capable of obtaining their associated scattering 
waves have been non-existing.
  An important step forward has recently been achieved with
  the release of the package {\small SIDES} to solve 
  Schr\"odinger's integro-differential equation.
With the introduction of {\small SWANLOP} package,
we provide an alternative broad-use tool to obtain scattering waves 
---and associated observables--- 
under any finite-range optical model potential,
regardless of its representation in coordinate- or momentum-space, 
or features in its nonlocality.

\section*{Acknowledgments}
\noindent H. F. A. is very grateful to colleagues of 
CEA,DAM,DIF for their kind hospitality during his stay at
Bruy\`eres-le-Ch\^atel, where part of this collaboration took place.
This research did not receive any specific grant from funding agencies
in the public, commercial, or not-for-profit sectors.

\appendix
\setcounter{table}{0}
\section{Relativistic corrections}
\label{sec:relativity}
Applications at high incident energies require the introduction of
for relativistic effects. 
Corrections of kinematical origin are incorporated as follows.
Let us consider a projectile of mass $m$ colliding a 
nucleus of mass $M$ at rest. 
The kinetic energy of the projectile in the laboratory reference 
frame is given by $E_L$.
Working in natural units $\hbar\!=\!c\!=\!1$, 
the projectile-target relative momentum $k$ in the 
center-of-momentum reference frame is given by
\begin{equation}
k^2 =
\frac{1}{4s}\left [s - (m+M)^2\right ]
	    \left [s - (m-M)^2\right ],
\label{k2rel}
\end{equation}
with the $s$-invariant given by $s\!=\!2ME_{L}\!+\!(m\!+\!M)^2$.
Additionally, the reduced mass $\mu$ needs to be replaced by
the reduced energy
\begin{equation}
\mu\to\frac{\varepsilon_p\,\varepsilon_t}{\varepsilon_p+\varepsilon_t}\,,
\label{rede}
\end{equation}
with 
$\varepsilon_p\!=\!\sqrt{k^2\!+\!m^2}$, and 
$\varepsilon_t\!=\!\sqrt{k^2\!+\!M^2}$.
The kinetic energy in the center-of-momentum reference frame is given by
$E\!=\!\varepsilon_p\!+\!\varepsilon_t\!-\!m\!-\!M$.
These corrections are obtained from Schr\"odinger's wave equation 
written in the center-of-momentum reference frame,
\begin{equation}
  \label{kinematics}
\left (
  \sqrt{m^2+{\bm p}^2} + \sqrt{M^2+{\bm p}^2} + U 
  \right )\Psi=
(\varepsilon_p+\varepsilon_t)\,\Psi\;,
\end{equation} 
followed by a first-order expansion of the square of
the relative momentum operator ${\bm p}^2$ around $k^2$.

\section{Coulomb potential}
\label{sec:coulomb}
The potential energy between a charged projectile (proton)
and the nucleus assumes a uniform proton density of radius $R_{_C}$.
Considering a target of charge $Ze$, then the potential energy
of the proton at a distance $r$ from the center of the nucleus is
given by
\begin{equation}
  \label{vcoul}
  V_{_C}(r) = 
  \left \{
  \begin{array}{ll}
  \displaystyle{
 \frac{Ze^2}{2R_{_C}}\left [3-2\left (\frac{r}{R_{_C}}\right )^2\right ]
    } & 
  \qquad\textrm{for $r<R_{_C}$;} \\
   & \\
  \displaystyle{\frac{Ze^2}{r}
    } &
    \qquad\textrm{for $r\geq R_{_C}$.}
  \end{array}
  \right .
\end{equation}
In the case of proton scattering using TPM parametrization
of PB nonlocal model, we adopt $R_{_C}\!=$\num{1.34}~fm. 
In all other cases we determine $R_{_C}$ using the extended liquid drop
model of Ref.~\cite{Myers1983}, where the charge root-mean-square
radius is parametryzed as
\begin{equation}
  \label{myers}
  \langle r^2 \rangle_{ch}^{\sfrac{1}{2}} =
  \sqrt{\textstyle{\frac{3}{5}}}\;A^{1/3}
  \left ( \num{1.15} + \num{1.80}\, A^{-2/3} - \num{1.20}\, A^{-4/3}
  \right )\;\textrm{fm} \;.
\end{equation}
To the resulting charge mean-squared-radius (m.s.r),
the proton charge m.s.r. $R_p^{\,2}$ is unfolded,
with $R_p\!=\!\num{0.875}$~fm~\cite{PDG2018}.
Therefore, the point-proton (pp) density m.s.r. becomes
\begin{equation}
  \label{unfold}
  \langle r^2 \rangle_{pp} = \langle r^2 \rangle_{ch} - R_p^{\,2} 
  = \textstyle{\frac{3}{5}}\,R_{_C}^{\,2}\;.
\end{equation}
From this expression we obtain $R_{_C}$ used by {\small SWANLOP}.
In the package, subroutine \texttt{vcoulomb.f} can be customized 
by the user to adapt alternative forms to calculate $V_{_C}(r)$.

\section{Integration entries and \texttt{LMAX}
under potentials read from file}
\label{sec:special}
When a local potential is read from file then 
\texttt{RMAX} and \texttt{NRP} are taken from that file,
while \texttt{LMAX} is defined by the user.
In the case of PB nonlocal potentials
(calculated internally by the code)
the values of \texttt{RMAX}, 
\texttt{NRP} and \texttt{LMAX} are fully controlled by the user.
However,
if a nonlocal potential is read from file all the above entries
are taken from that file.
In the case of
momentum-space potential read from file,
\texttt{LMAX} is taken from the number of angular points
over which the potential is defined,
while both \texttt{RMAX} and \texttt{NRP} are defined by the user.
All these considerations are summarized in Table~\ref{tab:special},
where checkmarks are placed on user-defined entries according
to \texttt{KPOT} definition.
  \noindent
   \begin{table} [H]
\begin{center}
    \begin{tabular} { c c c c }
 \hline
 \texttt{KPOT}& \texttt{RMAX} & \texttt{NRP} & \texttt{LMAX}\\
 \hline
            0 &     --        &     --       &     \ding{51}\\
            1 &     \ding{51} &     \ding{51}&     \ding{51}\\
            2 &     \ding{51} &     \ding{51}&     \ding{51}\\
            3 &     --        &     --       &     --       \\
            4 &     \ding{51} &     \ding{51}&     --       \\
 \hline
 \end{tabular}
\caption{ \label{tab:special}
  Checkmarks on integration entries and \texttt{LMAX} 
  to be specified by the user in 
  main input file depending on the potential choice \texttt{KPOT}.}
\end{center}
      \end{table}
      The code gives also the possibility of setting internally
      user-defined entries.
      To validate this action negative values must be supplied
      for the corresponding 
      \texttt{RMAX}, \texttt{NRP} and/or \texttt{LMAX}.
      In that case Eqs.~\eqref{rmax}, \eqref{lmax} and \eqref{hmax}
      are used, keeping $h\!\leq\!0.1$~fm, 
      with $R_{max}$ multiple of \num{0.5}~fm.

\section{Additional local potential option}
\label{sec:optionkadd}
Under setting \texttt{KADD=1} or \texttt{KADD=2} in line 11 
of \texttt{fort.1}, a local potential is added to the potential defined 
under \texttt{KPOT} option.
If \texttt{KADD=1}, the potential is read from \texttt{fort.22} 
by subroutine \texttt{read22.f},
replacing any existing hadronic local term.
If \texttt{KADD=2}, the additional local potential is calculated
by user-customized subroutine \texttt{user{\char`_}vloc.f}.
In Table~\ref{tab:kkadd} we summarize actions taken by 
{\small SWANLOP} under \texttt{KADD=0,1}, 
depending on \texttt{KPOT} value.
  \begin{table}[H]
    \begin{center}
      \begin{tabular}{r l}
 \fontsize{9}{12}\selectfont
        \texttt{KPOT} & Action \\
        \hline
        0 & Local term \textit{overwritten}\\
        1 & PB local term \textit{overwritten}\\
        2 & TPM local term \textit{overwritten} \\
        3 & Local potential \textit{superposed} \\
        4 & Local potential \textit{superposed} \\
        \hline
      \end{tabular}
      \caption{\label{tab:kkadd}
      Actions taken by {\small SWANLOP} under \texttt{KADD=1} or 2,
      depending on \texttt{KPOT} entry.
      }
    \end{center}
    \end{table}
\noindent

\section{Analytic scattering matrix for separable potential}
\label{sec:analytic}
In the absence of Coulomb forces, 
for a given rank-1 separable potential
$\hat V\!=\!|\xi\rangle\lambda\langle\xi|$,
the solution for the scattering matrix $\hat T(E)$ is given by
\begin{equation}
\label{tsol}
  \hat T(E) = \frac{|\xi\rangle\lambda\langle\xi|}
                 {1 - \lambda\langle\xi|\hat G_0^{(+)}(E)\,|\xi\rangle}\;,
\end{equation}
where $\hat G_0^{(+)}$ corresponds to the free propagator
for outgoing waves.
Projecting on-shell we get
\begin{equation}
  \label{te}
\langle k|\hat T(E)|k\rangle =
  t(E) = \frac{\lambda\,|\tilde\xi(k)|^2}
              {1 - \lambda\,\langle\xi|\hat G_0^{(+)}(E)\;|\xi\rangle}\;,
\end{equation}
with $E\!=\!\hbar^2 k^2/2\mu$, and
\begin{equation}
  \label{inner}
  \langle\xi|\hat G_0^{(+)}(E)\;|\xi\rangle =
  \frac{2}{\pi} \int_0^\infty 
  \frac{p^2dp\,|\tilde\xi(p)|^2}
       {E+i\epsilon-\hbar^2 p^2/2\mu}\;,
\end{equation}
where $\epsilon$ is a positive infinitesimal 
to account for outgoing waves.

In the case of the separable potential in Ref.~\cite{Bagchi1974}
\begin{equation}
  \label{urr}
  V(r',r) = \lambda \,\xi_n(r')\xi_n(r)\,,
\end{equation}
the form factors are defined in coordinate space given as
\begin{equation}
  \label{ffr}
  \xi_n(r) = \exp(-\alpha r)(\alpha r)^n/r\;,
\end{equation}
so that their corresponding form for $s$ waves 
in momentum representation becomes 
\begin{equation}
  \label{ffk}
  \tilde\xi_n(p)=\langle p|\xi\rangle =
  \int_0^\infty r^2 dr\,j_0(pr)\xi_n(r)\;.
\end{equation}

With the use symbolic manipulation software
{Mathematica\texttrademark} we obtain the explicit expressions
\begin{subequations}
  \label{xik}
\begin{align}
 \tilde\xi_1(p) &= \frac{2\alpha^2}{(\alpha^2 + p^2)^2} ; \\
 \tilde\xi_2(p) &= \frac{2\alpha^2 (3\alpha^2 - p^2)}{(\alpha^2 + p^2)^3}.
\end{align}
\end{subequations}
We apply these results to evaluate 
$X_n(k)\!\equiv\!\langle\xi_n|\hat G_0^{(+)}(E)\;|\xi_n\rangle$, 
in Eq.~\eqref{inner}, obtaining the closed-form  expressions
\begin{subequations}
  \label{xk}
\begin{align}
  X_1(x) &=-\frac{2\mu}{\hbar^2\alpha^3}
       \left [ \frac{(5 - 15 x^2 - 5 x^4 - x^6)} {4(1 + x^2)^4}
              + i\frac{4x}{(1 + x^2)^4} \right ]\;;\\
  X_2(x) &=-\frac{2\mu}{\hbar^2\alpha^3}
   \left [\frac{(11 -2 x^2 + 3 x^4)(3 - 19 x^2 -7 x^4 - x^6)}{4(1 + x^2)^6}
              + i\frac{4x(3 - x^2)^2}{(1 + x^2)^6} \right ]\, .
\end{align}
\end{subequations}
Thus, making use of Eqs.~\eqref{xik} and \eqref{xk},
the on-shell $T$ matrix in Eq.~\eqref{te} becomes
\begin{equation}
  \label{tee}
  t(E)=\frac{\lambda\; |\xi_n(k)|^2}{1 - \lambda\; X_n(k)}\;,
\end{equation}
where $k\!=\!\sqrt{2\mu E}$. 
In these units the $S$ matrix is expressed by
\begin{equation}
  \label{smatrix}
  S(E)=1 - 2i\,t(E)\;,
\end{equation}
to be used in Eqs.~\eqref{xsections} to evaluate cross sections.

In the case of form factors given by Eq.~\eqref{fr} we obtain
simple closed forms for the volume integral per nucleon 
of the potential,
  \begin{equation}
  \label{joa}
  J/A = \frac{1}{A} \int d^3r_1 \,d^3r_2 U(r_1,r_2)
  = \frac{\lambda}{A}
        \left [
          \frac{4\pi (n+1)!}{\alpha^2}
        \right ]^2\;,
      \end{equation}
and for the mean squared radius
\begin{equation}
\label{rms}
\langle r^2 \rangle = \frac{\int_0^\infty r^4 \xi_n(r) dr}
                           {\int_0^\infty r^2 \xi_n(r) dr}
                    = \frac{ (n+2)(n+3)}{\alpha^2} \;.
\end{equation}
These expressions become useful to calibrate values of $\alpha$
and $\lambda$, from estimates of bulk size of the 
targets and $J/A$ values.

\section*{References}
 \bibliographystyle{elsarticle-num}
%
%

%
%
  \end{document}